\documentclass[preprint]{aastex631}

\usepackage[utf8]{inputenc}
\usepackage{amsmath}
\usepackage{mathtools}   
\usepackage{graphicx}
\usepackage{xcolor}
\usepackage{verbatim}
\usepackage{comment}
\usepackage{changepage}

\newcommand*{\rom}[1]{\expandafter\@slowromancap\romannumeral #1@}

\def\wind{\textrm{wind}}
\def\NSC{\textrm{NSC}}
\def\BH{\textrm{BH}}

\begin{document}

\title{
        \Large
        The Inner 2 pc of Sagittarius A*: \\Simulations of the Circumnuclear Disk and Multiphase Gas Accretion in the Galactic Center
        }

\author[0000-0001-6541-734X]{Siddhant Solanki}
\affiliation{University of California, Santa Barbara, CA 93107, USA}
\affiliation{NASA Goddard Space Flight Center, 8800 Greenbelt Rd, Greenbelt, MD 20771, USA}
\affiliation{University of Maryland, 7901 Regents Drive, College Park, MD 20742, USA}

\author[0000-0003-0220-5723]{Sean M. Ressler}
\affiliation{Kavli Institute for Theoretical Physics, University of California Santa Barbara, Kohn Hall, Santa Barbara, CA 93107, USA}
\affiliation{Canadian Institute for Theoretical Astrophysics, 60 St. George Street, Toronto, ON M5S 3H8, Canada}

\author[0000-0001-8986-5403]{Lena Murchikova}
\affiliation{Black Hole Initiative, 20 Garden St, Harvard University, Cambridge, MA 02138, USA}
\affiliation{Institute for Advanced Study, 1 Einstein Drive, Princeton, NJ 08540, USA}
\affiliation{CIERA and Department of Physics \& Astronomy, Northwestern University, Evanston, IL 60208, USA}

\author{James M. Stone}
\affiliation{Institute for Advanced Study, 1 Einstein Drive, Princeton, NJ 08540, USA}

\author{Mark R. Morris}
\affiliation{University of California, Los Angeles, 430 Portola Plaza, Los Angeles, CA 90095, USA}

\begin{abstract} \label{abstract}
We present hydrodynamic simulations of the inner few parsecs of the Milky Way's Galactic Center that, for the first time, combine a realistic treatment of stellar winds and the circumnuclear disk as they interact with the gravitational potential of the nuclear star cluster and Sagittarius~A*. 
We observe a complex interaction of the stellar winds with the inner edge of the circumnuclear disk, which leads to the growth of instabilities, induced accretion of cool gas from the inner edge of the disk, and the eventual formation of a small accretion disk of $\sim 10^4-10^5$ K within $r \sim 0.1$ pc.
\end{abstract}

\section{Introduction} \label{Introduction}
\subsection{Structures in the Inner 5 Parsecs} \label{Introduction:Structures in the Inner 5 Parsec}

\begin{figure}[t]
    \centering
    \includegraphics[width=0.48\columnwidth]{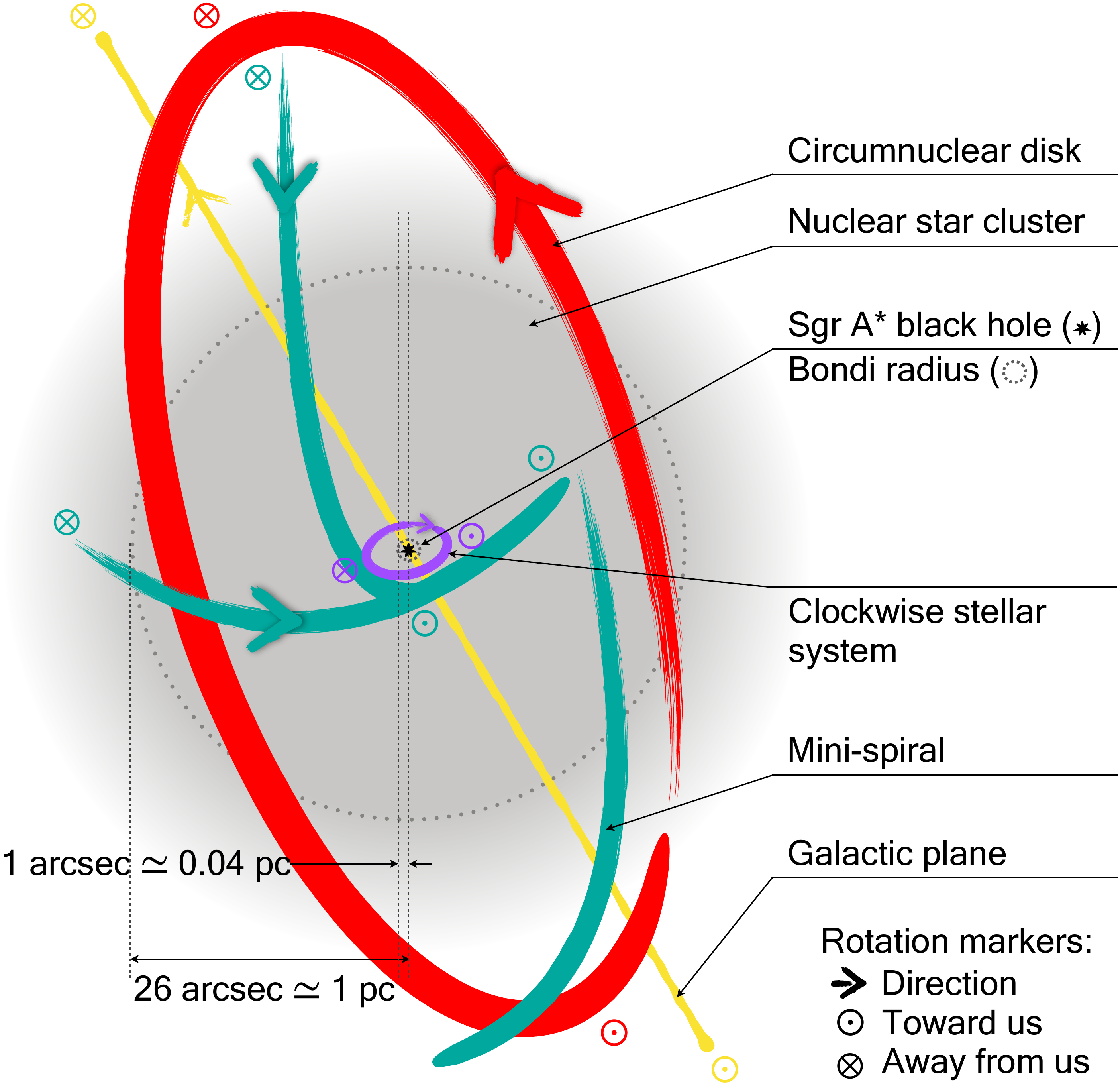}
    \caption{Schematic plot of the inner two pcs of the Galactic Center in the plane of the sky. The inner edge of the CND (red) is located at a distance of $1.4$ pc from the black hole. The Mini-Spiral streamers (turquoise) are present inside the cavity of the CND. Within $r=0.5$ pc is the clockwise stellar disk (purple) that contains the Wolf Rayet stars. The angular momentum of the disk about the line of sight is opposite to that of the CND.}
    \label{fig:schematic}
\end{figure}

The inner few parsecs of the Galactic Center is a complex region with many gaseous structures of various temperatures (Figure \ref{fig:schematic}). At the very center also lies the $4 \times 10^6~$M$_{\odot}$ black hole (BH), Sagittarius A* (Sgr A*), see \cite{Eckart_Genzel_1997MNRAS.284..576E,Ghez_1998ApJ...509..678G,Gravity2019,Do2019}. Orbiting this black hole is a Nuclear Star Cluster (NSC) that extends to several parsecs and has a mass of a few $10^6~$M$_{\odot}$  \citep{Genzel10}. There are over a million stars in the NSC within the central parsec where most of the observed stars are older giants with masses up to $4$ M$_\odot$ \citep{Genzel10,MWNSC_1,MWNSC_2}. The gravitational potential (and the mass distribution) of the NSC is inferred from the stellar orbits in the region \citep{Feldmeier_2014, 2017MNRAS.466.4040F, 2015MNRAS.447..948C}. Closest to the black hole is the group of ``S-stars'' which orbit the black hole within the central arcsecond, i.e. about 0.04 pc for the 8 kpc distance to the black hole \citep{ghez_s_stars, eckart_S_Stars}. The orbits of the S-stars have been crucial in determining the central black hole mass (e.g. \citealt{Gravity2019,Do2019}).  A small accretion disk has been reported based on the H30$\alpha$ line emission at $\sim 0.01$ pc that is primarily composed of $10^4$K gas \citep{Murchikova2019}. There are also several massive stars, many of which are post-main sequence (O-type, early B-type and Wolf-Rayet stars) that orbit within $0.5$ pc of the black hole \citep{paumard_massive_stars, lu_stars}. These stars have masses ranging from $20-120$ M$_{\odot}$ and about $20\%-50\%$ of them are rotating in a clockwise stellar disk as seen from Earth \citep{paumard_massive_stars, lu_stars, Levin_Beloborodov_2003ApJ...590L..33L, 2014ApJ...783..131Y} though recent studies suggest that this fraction is possibly higher \citep{2018ApJ...853L..24N}.  There is a sharp inner cutoff in the number of massive stars at $\sim 0.04$~pc and no massive O-type or Wolf-Rayet stars have been found inside this radius. The massive, young stars in the central $0.5$ pc, collectively called the young nuclear cluster (YNC), appear to have formed in a relatively recent event, 4 - 7 million years ago \citep{Krabbe__1995,2013ApJ...764..155L}. The Wolf-Rayet stars are associated with high mass loss rates, about $10^{-5} - 10^{-4}$ M$_{\odot}$ yr$^{-1}$ per star in the form of supersonic winds.

A dense ring of molecular and atomic gas called the Circumnuclear Disk (CND) orbits the central black hole in projection in a counter-clockwise fashion, that extends from $1.5$ to $3$ pc \citep{1982ApJ...258..135B,1989IAUS..136..393G}. Mass estimates for the CND vary between $10^4 - 10^6$ M$_{\odot}$ but a value of $3-4\times 10^4 \, \textrm{M}_{\odot}$ is generally accepted in the literature \citep{10.1093/mnras/stw771,Dinh_2021,James2021}.

The CND has a relatively evacuated central cavity with a $\sim 1$ parsec radius which contains several $100$ M$_{\odot}$ of neutral, partially ionized and fully ionized gas as well as molecular Hydrogen \citep{2016A&A...594A.113C}. The gas is ionized by radiation from the stars in the Young Nuclear Cluster \citep{Martins_stars}. This region in the center is referred to as the ``ionized-cavity'' or the ``Sgr A West HII region'', and hosts what is called the Mini-Spiral \citep{1983Natur.306..647L,lacy1991galactic_mini_spiral, Morris_GC_review}. The Mini-Spiral consists of a set of three streams of gas and dust in orbit around the central black hole \citep{Zhao2009,Tsuboi_2017_MS_orbits, 2012A&A...538A.127K_mini_spiral}. The total mass of the Mini-Spiral is estimated to be about $300$ M$_{\odot}$ \citep{jackson1993neutral}.  It has been previously argued that the Mini-Spiral gas filaments may have originated in the CND before falling in towards the black hole at a rate of about $10^{-3}\, \textrm{M}_\odot \, \textrm{yr}^{-1}$ \citep{Genzel_1994RPPh...57..417G}. 
Recent observations of the CND suggest a ``clumpy'' morphology \citep{Clumoy_disk_1_2001A&A...367...72V, Clumpy_disk_2_2002A&A...388..128V}. The clumps have reported (H$_2$) number densities and temperatures ranging between $10^3-10^8$ cm$^{-3}$ and $50$ and $500$ K \citep{Hsieh_2021}. Observations made with the HAWC+ instrument on the SOFIA observatory show that the inner few pcs of the GC contain strong magnetic fields - the CND shows toroidal fields that follow all the way to the Mini-Spiral \citep{Hsieh_2018}, Dowell et al. [preprint].

\subsection{Dynamics in the Galactic Center} \label{Introduction:Dynamics in the GC}

The diverse structures in the central $5$ pc interact with each other through gravity, radiation, and the exchange of matter. The black hole is fueled by in-falling matter from various scales \citep{2005ApJ...620..744G,Genzel10}. Recent works have shown that the winds from the Wolf-Rayet stars in the YNC can explain the accretion rates onto the central black hole and the strength of X-ray emission from the inner 2 pc \citep{1997ApJ...488L.149C,Cuadra:2007ba,10.1093/mnras/sty1146,Calderon_2020ApJ...888L...2C,Ressler_2020}. Realistic hydrodynamic and magnetohydrodynamic (MHD) simulations of Sgr~A* accretion were performed by \cite{Cuadra:2007ba,10.1093/mnras/sty1146,2020MNRAS.492.3272R}. They simulated the stellar winds from Wolf-Rayet stars and analysed the flow of matter towards the central black hole within about a $1$ pc radius. 
The observational constraints on the orbits and wind properties of the WR stars \citep{paumard_massive_stars, lu_stars} were used to inform simulations.
At scales of a few parsecs from the black hole the collective stellar winds result in a steady radial outflow for distances beyond $\gtrsim$ 0.5 pc.
In principle that collective wind should interact with the Mini-Spiral as well as with the CND, but neither of those structures were included in the aforementioned simulations.  It was believed until recently that the ram pressure from these winds is responsible for maintaining the central ionized cavity. The simulations of \cite{10.1093/mnras/stw771}, which modeled the WR stellar wind/CND interaction as a spherically symmetric wind surrounded by a thin, cold disk, showed that the winds instead accelerate the accretion from larger scales onto the black hole due to low angular momentum wind material getting deposited onto the CND. This accretion from the CND results in an inflow of relatively cold material, in contrast to the hot stellar winds from the Wolf-Rayet stars. 

\vspace{10pt}

Multiphase accretion in the central few parsecs of the Galactic Center has not been systematically studied. Observed gaseous structures in the inner $0.01 - 2$ pc have been modelled predominantly independently of each other. In this work we track the combined accretion of the CND and the stellar winds on $0.01-1$ pc scales using 3D, hydrodynamic simulations. 
In this way we bridge the gap in scales between \citet{2016MNRAS.459.1721B} and \citet{10.1093/mnras/sty1146} to form a more unified model of this region.
The result is the most detailed and the closest to observations model of the inner few parsecs of the Galactic Center.

\vspace{10pt}

This paper is organized as follows. In Section \ref{Methods} we discuss the methods we adopt in constructing the simulation. In Section \ref{Results} we explain the main results from our analysis of the simulation. We discuss our results in Section \ref{Discussion} as well as talk about the limitations of the current models of the CND. We summarize our primary conclusions in Section \ref{Conclusions}.

\section{Methods} \label{Methods}
We use the multi-purpose fluid dynamics code {\tt Athena++} \citep{Stone_2020}. {\tt Athena++} is an astrophysical magnetohydrodynamics code that is used to perform simulations of astrophysical fluids and plasmas. It solves the conservative equations of fluid mechanics using finite volume methods.
We adopt the Harten-Lax-van Leer-Einfeldt (HLLE: \citealt{Einfeldt1988}) Riemann solver and piecewise-linear method reconstruction.

\begin{figure}
\centering
\begin{tabular}{cc}
\includegraphics[width=0.45\textwidth]{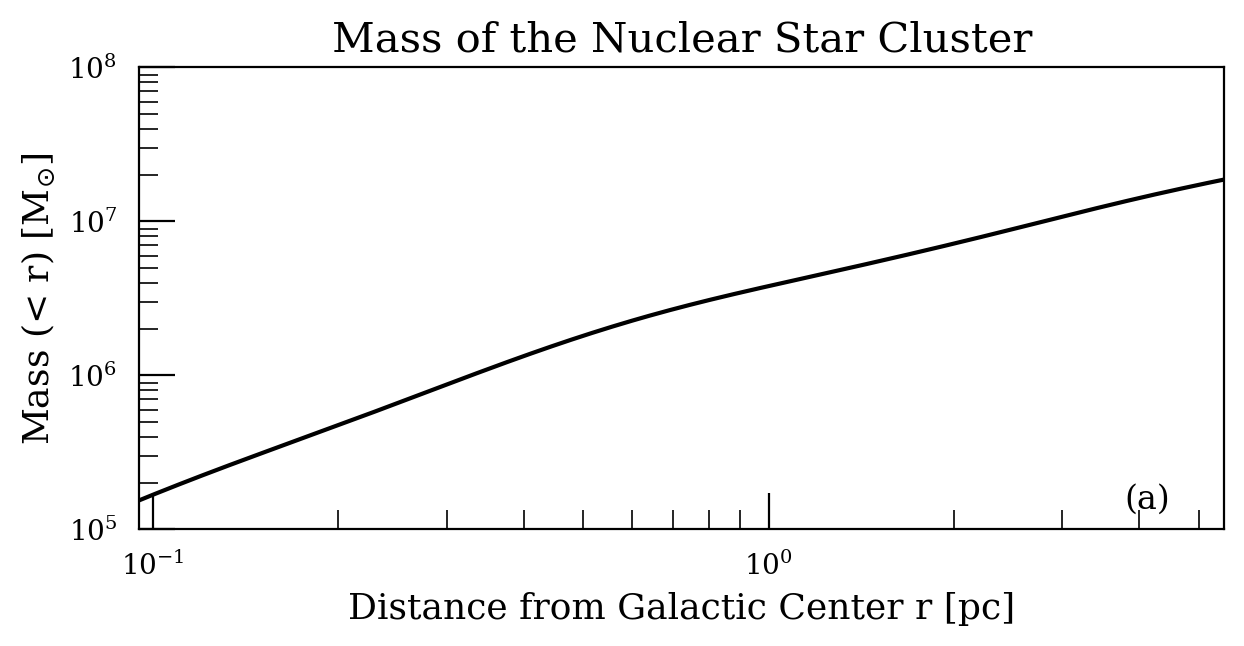} &
\includegraphics[width=0.45\textwidth]{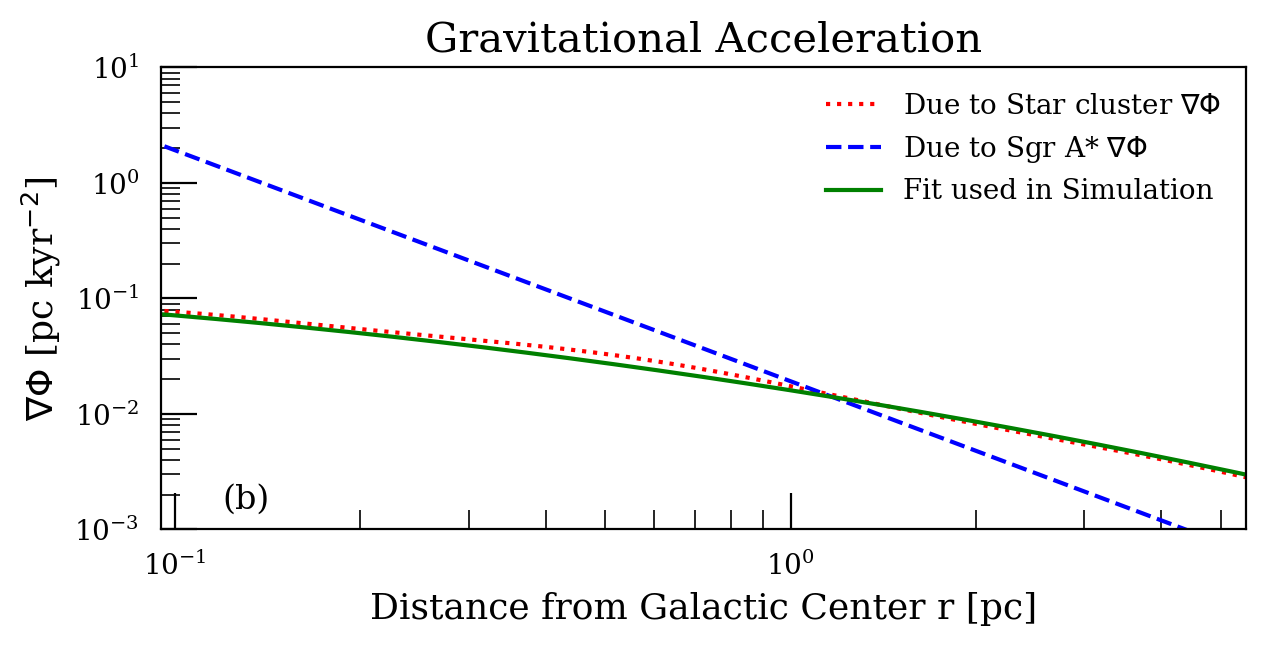}
\end{tabular}
\caption{Properties of the Nuclear Star Cluster (NSC) in the Milky Way's Galactic Center. Left: Cumulative mass profile of the NSC, i.e. combined mass of the NSC's stars enclosed within radius $r$ from the Galactic Center as a function of $r$ (see also \citealt{2017MNRAS.466.4040F}).  Right: Contribution of NSC (red) and the Supermassive black hole (blue) to gravitational acceleration in the inner 5 pc of the Galactic Center as a function of $r$. The range of radii include all of our simulation domain. We used an analytical fit (green) to account for the potential of the NSC. We then add a $\frac{GM_{\textrm{BH}}}{r^2}$ term to the gradient of the NSC potential fit to account for the total gravitational acceleration.
}
\label{fig:NSC_potential}
\end{figure}

To the public version of \texttt{Athena++}, we have added the effects of optically thin cooling and stellar winds. We solve the following set of non-relativistic hydrodynamic equations in conservative form: 
\begin{align}
    \frac{\partial \rho}{\partial t} + \nabla \cdot (\rho \textbf{v}) &= f \dot{\rho}_{\wind} \label{rho_cons}\\
    \frac{\partial (\rho \textbf{v})}{\partial t} + \nabla \cdot (P \textbf{I} + \rho \textbf{v} \textbf{v}) &= - \frac{\rho G\textrm{M}_{\BH}}{r^2}\hat{r} - \rho \nabla \Phi_{\NSC} + f \dot{\rho}_{\wind} \left<\textbf{v}_{\text{wind, net}}\right> \label{mom_cons}\\
    \frac{\partial E}{\partial t} + \nabla \cdot ((P + E) \textbf{v}) &= - \frac{\rho G\textrm{M}_{\BH}}{r} \textbf{v} \cdot \hat{r} - \rho \textbf{v} \cdot \nabla \Phi_{\NSC} + \frac{1}{2} f \dot{\rho}_{\wind} \left<|\textbf{v}_{\text{wind, net}}|^2\right> - \mathcal{Q_{-}} \label{eng_cons}
\end{align}
Here $\rho$ is the fluid density, $\textbf{v}$ is the fluid velocity vector, $P\textbf{I}$ is the diagonal thermal pressure tensor, $E$ is the fluid energy density, $\Phi_{\NSC}$ is the Nuclear Star Cluster's gravitational potential, $G\textrm{M}_{\BH}$ is scalar gravitational parameter of Sgr A*, $\mathcal{Q_{-}}$ is the scalar function describing cooling rate per unit volume (\ref{Methods:Cooling}), $\hat{r}$ is the unit position vector from Sgr A*, and $\dot{\rho}_{\wind}$ and $\mathrm{v}_{wind}$ are the stellar wind mass and velocity source terms in the frame of the grid (see paragraph below). $f$ is the fractional volume of a cell occupied by the stellar wind and $ \textbf{v} \textbf{v}$ is the vector product of the fluid velocity with itself.

These equations represent the conservation of mass (Equation \ref{rho_cons}), conservation of momentum (Equation \ref{mom_cons}) and conservation of energy (Equation \ref{eng_cons}) per unit volume respectively. The right hand sides of the above equations include source terms from stellar winds that represent additional mass, momentum and energy that is being added or removed from the system. The simulation is performed on a Cartesian grid with static mesh refinement (see \ref{Methods:Grid}). There are $\sim$ $30$ stars that move in fixed Keplerian orbits. In the frame of each ``star'', a stellar wind emanates spherically with a feeding radius of $r_{\wind}$, equal to twice the length of the diagonal of the mesh-refined cell at which the star is located. Inside of this sphere there is a constant source of mass that is determined from the observed mass loss rate of that particular wind, $\dot{M}_{\wind}:\dot{\rho}_{\wind} = \dot{M}_{\wind}/V_{\wind}$, where $V_{\wind} = 4\pi/3 r_{\wind}^3$ is the volume of the wind. To compute $f$ and $\left<...\right>$, each cell that intercepts the sphere at which a star is centered is broken down into a $5\times5\times5$ sub-grid over which an average is performed.

We include a point-mass gravitational potential with a mass of M$_{BH} = 4.2 \times 10^6 \text{M$_{\odot}$}$ to capture the effects of Sgr A* \citep{Gravity2019}. If we use the value of $3.98 \times 10^{6}$ M$_{\odot}$ obtained by \citealt{doi:10.1126/science.aav8137}, it would introduce only about 5\% difference and would not impact the dynamics at qualitative level. In addition to the black hole's gravity, we include the gravitational potential from the nuclear star cluster given by \citet{2017MNRAS.466.4040F} (Figure \ref{fig:NSC_potential}). We ignore the anisotropy in the potential inferred from the velocity dispersion of the stars in the NSC (\citealt{2017MNRAS.466.4040F} find $|{\sigma_z}/{\sigma_r}| \sim 0.9$ at $1$ pc, where $\sigma_z$ and $\sigma_r$ are the radial and plane of the sky velocity dispersions respectively), and assume spherical symmetry of the NSC mass distribution. To speed up the computation, we fit the following function to the NSC gravitational acceleration in the range of radii between $10^{-3}$ and $10^{1}$ pc:
\begin{equation}
    \ln (\nabla \Phi) = a \ln(r)^2 + b \ln(r) + c,
    \label{eq:NSC Phi}
\end{equation}
where our best fit parameters are $a = -0.0843 \mbox{ pc kyr$^{-2}$} , b = -0.8428 \mbox{ pc kyr$^{-2}$} \mbox{ and } c = -4.1338$ \mbox{ pc kyr$^{-2}$}. Here $r$ is taken in parsecs. As evident from Figure \ref{fig:NSC_potential} the fit is nearly identical to the observationally derived values.

We ran simulations for $450 \times 10^3$ years, which is roughly equal to the lifetime of the Wolf-Rayet stage of stellar evolution \citep{WR_lifetime}.

We outline our treatment of Wolf-Rayet stellar winds in Section \ref{Methods:Wolf-Rayet Stars}. There, we describe the set-up of a CND model disk in hydrodynamic equilibrium in Section \ref{Methods:Initializing the CND}, the cooling function employed in our simulations in Section \ref{Methods:Cooling}, the grid on which the simulation was performed in Section \ref{Methods:Grid}, and the simulation floors and boundaries in Section \ref{Methods:Simulation Floors}.

\subsection{Wolf-Rayet Stars} \label{Methods:Wolf-Rayet Stars}
We treat the winds of the $\sim$ $30$ Wolf-Rayet stars observed in the Galactic Center as source terms in mass, energy and momentum, 
following \cite{10.1093/mnras/sty1146}. The stars are initialized on fixed Keplerian orbits based on their current positions. The  Cartesian positions, velocities, and orbital parameter of the stars are observationally determined (Table 2 in \citealt{paumard_massive_stars}). For some of the stars the radial distances are unknown because their acceleration measurements are consistent with zero.

The orbits of these stars are then determined by choosing a $z$ position (radial) distance such that the eccentricity of the orbit is minimized, following  \citet{Cuadra:2007ba} and \citet{10.1093/mnras/sty1146}. Since we evolve the WR stellar winds along their orbits for 450 kyrs or $\sim$ $10-100$ orbital periods, the initial orbital phase of the stars is relatively unimportant.

\begin{figure}
    \centering
    \includegraphics[width=0.6\textwidth]{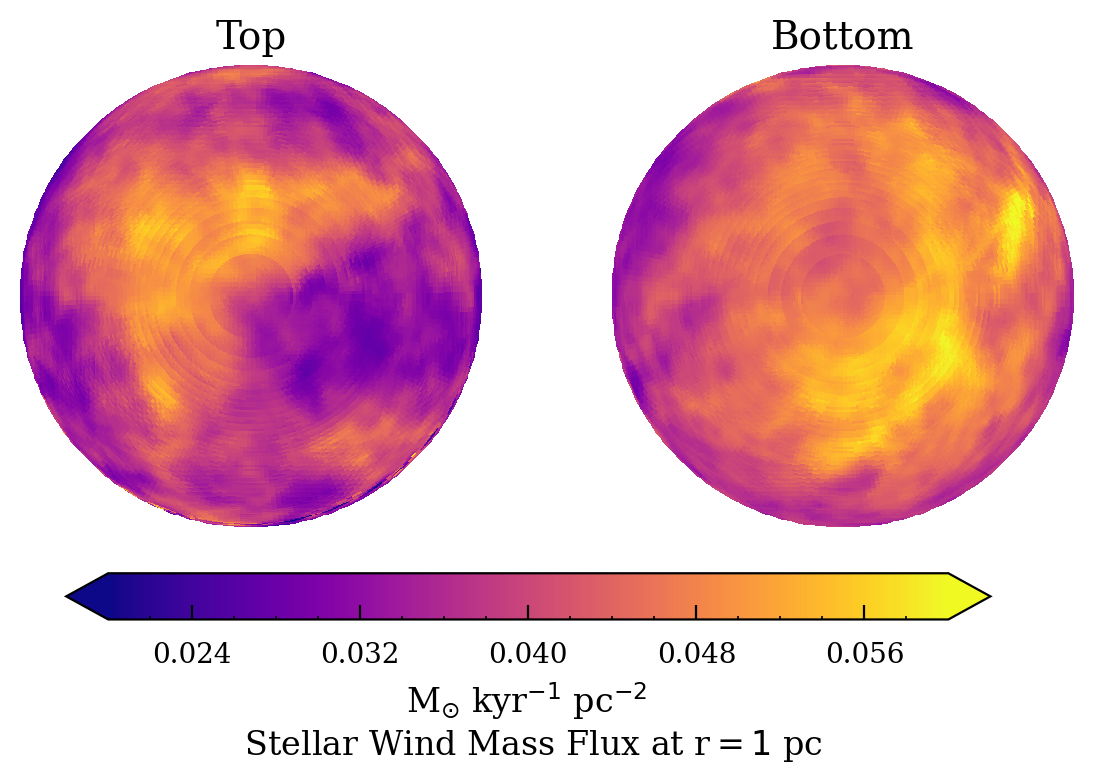}
    \caption{Time-averaged stellar wind mass flux from $10-100$ kyr projected onto a hemispheres of radius $1$ pc. We present the top and bottom view of the sphere. The anisotropies that originate from from the distribution of the stars are within a factor of $3$.}
    \label{fig:time_averaged_flux_per_area}
\end{figure}

To visualize the distribution of the winds over time, in Figure \ref{fig:time_averaged_flux_per_area} we plot the time-averaged radial mass flux (mass times velocity per unit area) of the stellar winds. The flux is averaged over $10-100$ kyr and projected onto a sphere of radius $1$ pc. The average is performed after $10$ kyr to allow the stellar winds to first reach the outer boundary of the simulation. We show that over this timescale, the anisotropies that originate from from the distribution of the stars are within a factor of $3$ and the combined winds approach an isotropic flow. This means that the inner edge of the CND, which is located at $1.5$ pc, receives, on average, an approximately uniform flow from the winds.

This uniform flow approaches a Parker wind solution \citep{parker_wind}, with the mass flux scaling as $(r^2 \rho v) = \mathrm{const}$.

\subsection{Initializing the CND} \label{Methods:Initializing the CND}
We use an isentropic (i.e. constant entropy) torus as a model for the CND. This is a common choice of a closure relation between pressure and density while initializing torii in, for instance, the general relativistic MHD community \citep{GRMHD_torii_1, GRMHD_torii_2}. 
The CND has been reported to have a clumpy, not a smooth distribution \citep{Clumoy_disk_1_2001A&A...367...72V, Clumpy_disk_2_2002A&A...388..128V}. Resolving such clumps in a simulation, however, requires significantly higher resolution than we can afford. We posit that a smooth disk model is sufficient to explain the bulk accretion properties of the CND, though we discuss its limitations in \ref{Discussion:Limitation of the Model}.  We note that previous authors have also modelled the CND as a smooth disk (e.g., \citealt{10.1093/mnras/stw771}).
For such a disk, the vanishing velocity time derivative implies a steady state,
\begin{equation}
    \frac{\partial \textbf{v}}{\partial t} = -\frac{\nabla P}{\rho} - \nabla \Phi - \textbf{v} \cdot \nabla \textbf{v} = 0,
    \label{eq:hydrodynamic_eq}
\end{equation}
where $\textbf{v}$ is the fluid velocity, $P$ is the fluid pressure, $\rho$ is the fluid pressure and density, and $\Phi$ is the gravitational potential. 
The velocity profile for the disk is purely azimuthal, that is, in the $\hat{\phi}$ direction, and depends only on the cylindrical radius $R$ from the center. We set the angular velocity to be $\Omega(R) \propto R^{-q}$, where $q=1.2$ and the radius of the pressure maximum is $R_{k} = 1.8$ pc from the black hole. These two parameters determine the thickness and radial extent of the disk.\footnote{For example, a perfectly Keplerian flow with $q=1.5$ and a point source gravitational potential would result in an infinitely thin disk.} We choose the particular values for $q$ and $R_k$ to get an overall mass of $3 \times 10^4$ M$_{\odot}$. \citep{Tsuboi_2017_MS_orbits,James2021}. The velocity is then:

\begin{equation}
    \textbf{v} = R \Omega(R) \hat{\phi} = CR^{-q+1} \hat{\phi} = \left(R_{k}^q \sqrt{\nabla \phi({R_k}) R_{k} }\right) R^{-q+1} \hat{\phi}  \label{velocity_profile}
\end{equation}
Further, for an isentropic flow, the pressure and density are related as $P = K \rho^{\gamma}$, where $\gamma$ is the adiabatic index, equal to $5/3$ for a monoatomic gas and $K$ is a constant factor set from the initial conditions. The equilibrium solution for the disk is scale-free in mass so we still have a remaining degree of freedom. We remove this freedom by appealing to observations. \citet{Hsieh_2021} and \citet{James2021} report that the CND have number densities of around $10^5$ cm$^{-3}$. Thus in our model we assume the number density to be  $10^5$ cm$^{-3}$ ($\sim 5000$ M$_{\odot}$ pc$^{-3}$) at the pressure maximum.

One must know the mean molecular weight, $\mu$, to compute the number density, $n$, from the mass density, $\rho$, via $n = \rho/\mu$. $\mu$ is taken to be equal to $1.39 m_p$. We discuss this choice of $\mu$ in Section \ref{Methods:Cooling}. In terms of composition, The CND consists of mostly molecular hydrogen but is also composed of heavier molecules. For instance, rotational lines from HCN, HCO$^+$ and CS molecules have been used as tracers of the CND (\cite{2001ApJ...551..254W, Hsieh_2021}). 

In order to get the observed thickness of the CND, which is about $0.2$ pc at the inner edge \citep{10.1093/mnras/stw771}, we use an initial temperature at the pressure maximum of $4500$ K, which is also the temperature floor. This is a factor of $10$ higher than the observed temperatures \citep{Tsuboi2017}. However, the pressure support in the CND comes from microturbulence and not thermal pressure, the former being equivalent to thermal pressure corresponding to a temperature of $10^4$K. Therefore our temperature can be thought of as a measure of the internal velocity dispersion that gives rise to microturbulence instead of the true temperature of the disk. Moreover, our choice of temperature gives us a reasonable dimension and aspect ratio, $H/r$ (where $H$ is the scale height), for the disk while maintaining the hydrodynamic equilibrium condition. A lower temperature floor causes the disk to cool to lower temperatures, losing the pressure support that maintains the disk thickness.  The temperature for the disk is still much lower than the temperature of shocked stellar wind material in the simulation ($T_{wind} \geq 10^6$ K) and is not dynamically important. In fact, it is the geometry of the disk that is more relevant to the overall evolution as we show in \ref{Discussion:Angular Momentum Transport in the Disk}. Our parameters are otherwise similar to \citet{10.1093/mnras/stw771}. It follows then that:

\begin{align}
    \frac{\nabla P}{\rho} &= \nabla \left(\frac{\gamma K P}{(\gamma - 1)\rho} \label{press_grad} \right) \\
    P &= \frac{\rho k_b T}{\mu} = K \rho^\gamma \\
    \longrightarrow K &= \frac{k_b T_{max}}{\rho_{max}^{\gamma-1}}
    \label{K_fac} 
\end{align}

Eq. \ref{eq:hydrodynamic_eq} can then be rewritten as a gradient and integrated:

\begin{equation}
     -\frac{\nabla P}{\rho} - \nabla \Phi - \textbf{v} \cdot \nabla \textbf{v} = 
     \nabla \left(\frac{\gamma K^2 \rho^{\gamma-1}}{(\gamma - 1)} - \Phi - \frac{C^2 (-q+1)R^{-2q+2}}{-2q+2}\right) = 0
     \label{eq:integrated_eq}
\end{equation}

The density, momentum and energy density are subsequently calculated from the pressure using the isentropic assumption.

\subsection{Cooling} \label{Methods:Cooling}

In reality, the composition of gas in the region of interest in the Galactic Center is complicated. WR stellar winds tend to be hydrogen-deficient and shocked to high temperatures $(T \geq 10^6 K)$ at which they become fully ionized. Gas in the CND, however, is mostly hydrogen and gets progressively more ionized near the ionized central cavity. The temperature of the gas in the Galactic Center ranges from about $10^2 \, \mathrm{K}$ to about $10^9 \, \mathrm{K}$, and it is necessary to have an accurate cooling function across this range. To correctly account for the gas cooling locally through various processes (a) the local gas composition must be known and tracked; and (b) a composition-dependent cooling rate must be computed and applied everywhere in the simulation. This procedure is generally not adopted because of computational costs and uncertainties in the composition. Instead, a global, temperature-dependent cooling function is parameterized using the gas temperature, which can be computed from the pressure, density and a mean molecular weight (see also \citealt{2017ApJ...846..133K}, \citealt{Koyama_2001}). With a cooling function $\Lambda(T)$ and a number density $n$, the cooling rate in $\mbox{erg s$^{-1}$ cm$^{-3}$}$ is given by $n^2\Lambda(T)$. We adopt a mean molecular weight of $\mu=1.39$ $m_p$ for the entire simulation and use two different cooling functions [$\Lambda(T)$] for low ($T < 10^{4.2}$ K $\sim 16000$ K) and high ($10^{4.2} \, \textrm{K} \leq T \leq 10^9 \, \textrm{K}$) temperature gas respectively. We follow \cite{Ressler_2020} for the stellar wind composition of hydrogen-deficient, fully ionized gas at 3 times solar metallicity, which gives $\mu = 1.39 m_p$. While the CND is known to contain molecular gas and therefore have a slightly different mean molecular weight, we set it to be $1.39$ m$_{p}$ for simplicity. 
Note that this is only a few percent different from the mean molecular weight of interstellar medium gas at solar metallicity ($\mu = 1.295$).
As the composition of the CND is uncertain, using something close to solar metallicity is a reasonable first-order assumption.

\begin{figure}
    \centering
    \includegraphics[width=0.95\textwidth]{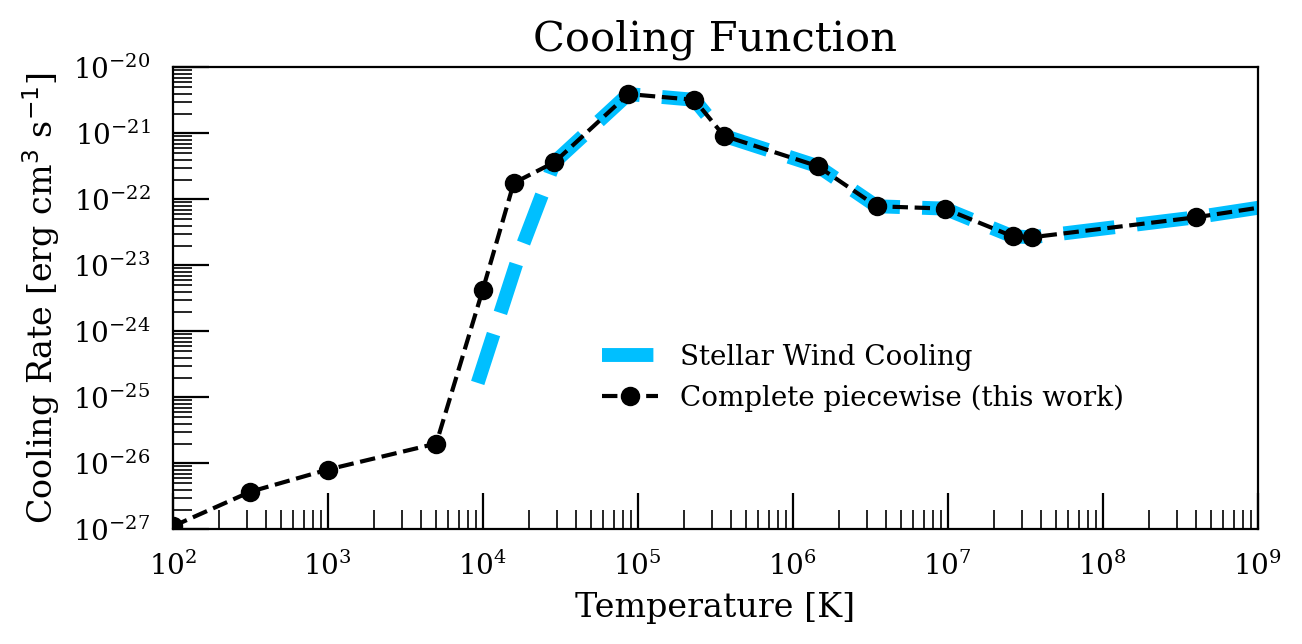}
    \caption{The cooling curve (black dashed and dotted line) used in the simulation.
    The dashed blue curve represents the cooling curve used by \citet{10.1093/mnras/sty1146} that we extended to lower temperatures using \citet{Koyama_2001}. The higher temperature portion ($\gtrsim 10^4 K$) is appropriate for completely ionized gas while the lower temperature portion ($\lesssim 10^4 K$) is appropriate for the cold, largely molecular gas. We approximate the curves using a piece-wise power-law fit.
    }
    \label{fig:cooling_curve}
\end{figure}

We combine the low temperature cooling function appropriate for the cold interstellar medium described in \cite{Koyama_2001} with the high temperature cooling function appropriate for the hot gas described in \citet{10.1093/mnras/sty1146} (see also, \citealt{Martins_stars} and \citealt{Cuadra:2007ba}). The low-temperature part accounts for effects of atomic line and reverberational cooling along with molecular collisions with grains. The high-temperature part accounts for effects of line emission in collisional ionization equilibrium and thermal bremsstrahlung.  The cooling function is shown in Figure \ref{fig:cooling_curve}.  

Note that for the purposes of calculating the cooling rates, we assume neutral, solar composition gas at temperatures below $10^{4.2}$K, and assume fully ionized, hydrogen-deficient, 3 $\times$ solar metallicity gas for higher temperatures. It is only for computing $n$ from $\rho$ that we always use $\mu=1.39m_{\rm p}$.

The actual integration of the cooling functions into the code is done using an exact analytical scheme proposed by \cite{Townsend_2009} for a piecewise power law fit to the cooling functions.

\subsection{Grid} \label{Methods:Grid}
The simulations are performed in a Cartesian box with dimensions of $10 \times 10 \times 10$ pc  along the $x$, $y$, and $z$ axes respectively, where $z$ aligns with the angular momentum of the disk as determined from observations. This direction is tilted with respect to the plane of the sky by $66$ degrees. The $x$ and $y$ axes are chosen such that they form the semi-minor and semi-major axes respectively, of the CND in the plane of the sky and the $y$ axis points north-eastwards. We particularly focus resolution within a cylinder centered on the black hole of height 1 pc and radius 2 pc using static mesh refinement. 

The grid has a base resolution of $128 \times 128 \times 128$ cells and 5 levels of static mesh refinement. The coordinate limits of each level of refinement are given in Table \ref{tab:refinement}.  On each successive level the cell resolution is increased by a factor of two.

\begin{deluxetable}{ccccc}
\tabletypesize{}
\tablecaption{Levels of mesh refinement boundaries} 
\tablewidth{0pt}
\tablehead{
\colhead{Level } & \colhead{$x_\mathrm{min}=y_\mathrm{min}$} &
\colhead{$z_\mathrm{min}$} & \colhead{$x_\mathrm{max}=y_\mathrm{max}$} &
\colhead{$z_\mathrm{max}$} \\
\colhead{number} & {pc} & {pc} & {pc} & {pc}
}
\startdata
0 & -5 & -5 & 5 & 5\\
1 & -3 & -0.7 & 3 & 0.7 \\
2 & -2.8 & -0.5 & 2.8 & 0.5\\
3 & -0.625 & -0.625 & 0.625 & 0.625\\
4 & -0.3125 & -0.3125 & 0.3125 & 0.3125\\
5 & -0.15625 & -0.15625 & 0.15625 & 0.15625
\enddata
\tablecomments{Coordinate boundaries for static mesh refinement in our simulations. The levels are chosen in a manner such that the resolution at the disk-wind interface is at least $0.02$ pc.}
\label{tab:refinement}
\end{deluxetable}

The outer boundary is located at $\pm 5$ pc for each of the $x, y$ and $z$ axes. We adopt an ``outflow'' boundary condition where matter that leaves the outer boundary is completely removed from the simulation domain. To prevent accumulation of matter at the center, the matter that enters a central sphere of radius two times the width of the smallest cell ($0.002$ pc in our simulations) is replaced with pressure and density floors with zero velocity. We follow the treatment from \citet{10.1093/mnras/sty1146}. The net accretion rate through the inner boundary of our simulations should not be taken as a prediction for the net accretion rate at the event horizon of the black hole, which is located at much smaller radii.  The artificially large inner boundary likely causes the net accretion rate at the inner boundary to be artificially large in a way typical of flows with a rough balance between inflow/outflow (see Appendix B in \citealt{Ressler_2020}).

\subsection{Simulation Floors and Boundaries} \label{Methods:Simulation Floors}

The primitive variables of density and pressure can occasionally fall to negative values as \texttt{Athena++} evolves the conservative values. In order to prevent that from happening, we set pressure and density floors of $10^{-10}$ M$_{\odot}$ pc$^{-1}$ kyr$^{-2}$ ($6.5 \times 10^{-17}$ erg cm$^{-3}$) and $10^{-7}$ M$_{\odot}$ pc$^{-3}$ ($6.8 \times 10^{-30}$ g cm$^{-3}$), respectively.  Whenever the pressure and density values fall below these floors, they are replaced by the floor values. To prevent the gas from over-cooling and letting the temperature (pressure) of the gas fall to a negative value, we also set up a temperature floor of $4500$ K, that acts as a density-dependent pressure floor.

Finally, since the the fluid sound speed sets the size of the time-step in the simulation, we impose a velocity ceiling of $5$ pc kyr$^{-1}$ ($4900$ km s$^{-1}$), which is equal to the free fall velocity at $0.004$ pc from the black hole (that is also the inner boundary). 
This limit is rarely used except for handful of cells with extremely large velocities in transient phases of the simulation (e.g., when the wind source terms are first initialized).   

\section{Results} \label{Results}

In the following subsections we describe the results from our analysis of the simulations. In Section \ref{Results:Evolution of the Disk} we summarize the general evolution of the disk in the presence of the stellar winds. Section \ref{Results:Inflow and Outflow} describes the inflow and outflow rates of the gas at various radii and times in the simulation. Section \ref{Results:Inner Edge Instabilities} details the inner edge instabilities in the disk. In Section \ref{Results:High and Low Temperature Flows} we partition the gas in the simulation into two families based on temperature and present their respective flow properties. Finally, in section \ref{res:small_accretion_disk} we analyze an accretion disk that forms at smaller radii later in the simulation.

\subsection{Evolution of the Disk} \label{Results:Evolution of the Disk}

\begin{figure}
    \centering
    \includegraphics[scale=0.48]{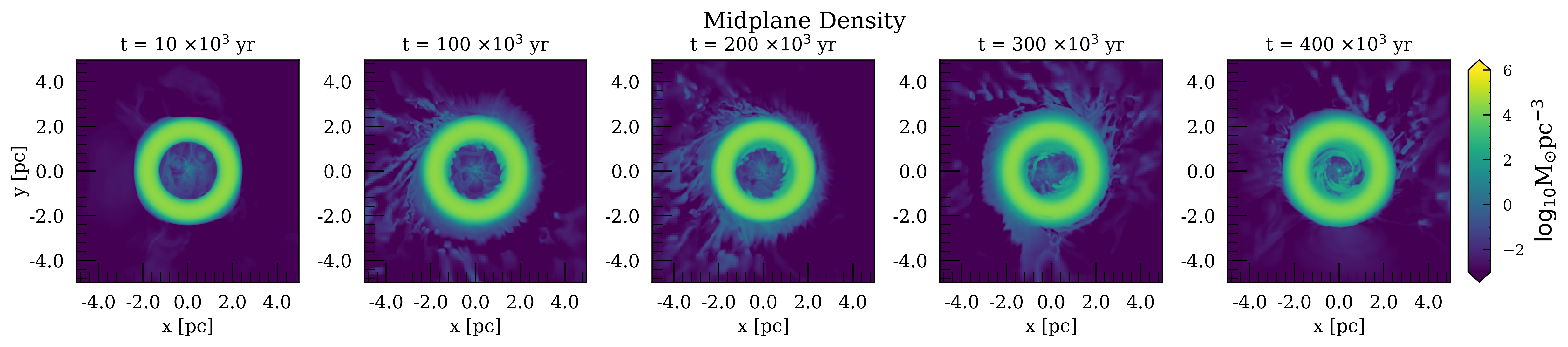}
    \includegraphics[scale=0.48]{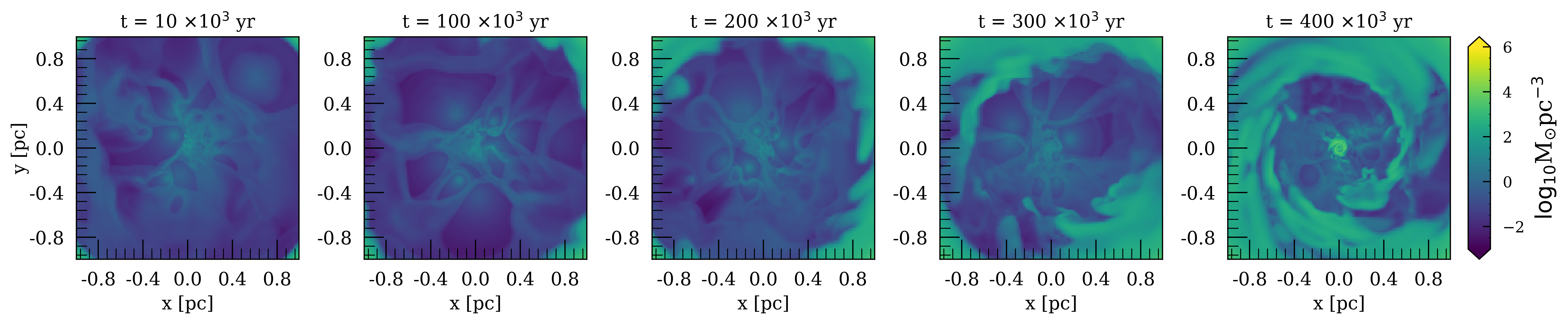}
    \caption{Midplane slices of density taken at different times in the simulation. The top panel shows the entire simulation domain and the bottom panel zooms in on a $2\times 2$ pc scale. The stellar winds, which are more visible in the bottom panel, cause inward migration of gas from the inner edge of the disk as they deposit mass. Mass from the outer edge of the disk moves outwards because of turbulent angular momentum transport within the disk. At late times (bottom right panel) the accreting gas from the CND forms a smaller accretion disk at $r \sim 0.1$ pc.}
    \label{fig:midplane_density}
\end{figure}

\begin{figure}
    \centering
    \includegraphics[width=0.95\textwidth]{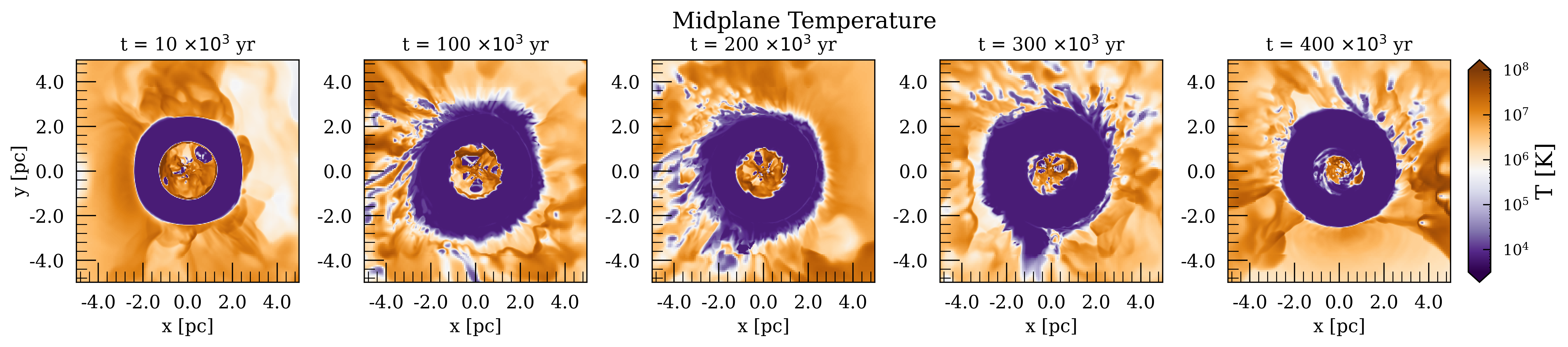}
    \includegraphics[width=0.95\textwidth]{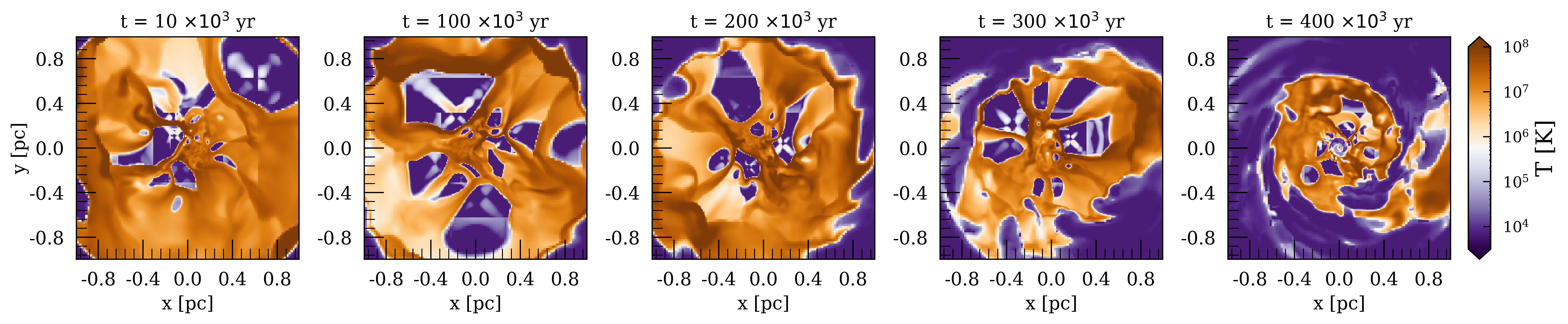}
    \caption{Midplane slices of temperature on the same scale as Figure \ref{fig:midplane_density}. The main body of the disk maintains its initial temperature of $T=4.5 \times 10^3$ K while the shocked gas from the stellar winds tends to have $T\gtrsim10^6$ K. 
    There is very little gas present in the $10^5 - 10^6$ K region.
    As in Figure \ref{fig:midplane_density}, cold gas is expelled near the outer edge of the disk due to outwards angular momentum transport.  
    Note that the oddly shaped low temperature (pressure) regions in the bottom panel are an artifact of the low number of cells within the feeding radius of the stellar winds. As described in Section \ref{Results:Evolution of the Disk}, these regions are dominated by ram pressure where the thermal energy represents a negligible fraction of the local energy budget.}
    \label{fig:midplane_temp}
\end{figure}

We plot the midplane density and temperature of the simulation in Figures \ref{fig:midplane_density} and \ref{fig:midplane_temp} at $5$ different times $(t = 10, 100, 200,  300 \mbox{ and } 400 \mbox{ kyr})$ in the simulation. 
The top panels in both the figures show the entire $5\times5$ pc simulation domain, whereas the bottom panels focus on $2\times2$ pc scales. 
The bulk of the disk maintains its initial temperature of $4.5 \times 10^3$ (the temperature floor of the simulation) and is $\sim 4$ orders of magnitude denser and $\sim$ 4 orders of magnitude colder than the collective material from the stellar winds.
The disk traps most of the winds within its inner edge in the midplane, causing a shock of width $\approx 0.1$ pc at the inner edge of the disk that is apparent in the temperature plot. The interaction between the winds and the disk also leads to the formation of “cartwheel” instabilities that transport angular momentum outwards and drive inflow to smaller radii \citep{10.1093/mnras/stw771}. These instabilities have the characteristic length scale of a few tenths of a pc and form within about $100$ kyrs. 
The top panels show cold filamentary structures at the outer edge of the disk that carry mass and angular momentum outwards away from the disk. 
The bottom panel of Figures \ref{fig:midplane_density} and \ref{fig:midplane_temp} show the individual stellar winds more clearly as relatively high density, low temperature regions surrounded by high density, higher temperature bow shocks. Throughout the simulation, there is only a small amount of gas present in the intermediate temperatures between the wind-shocked gas and the disk ($10^5 \lesssim \mbox{T} \lesssim 10^6$. This enables us to use temperature as a tracer to differentiate between the stellar wind and disk material (as we do in Section \ref{Results:High and Low Temperature Flows}).

Visible in the bottom panel of Figure \ref{fig:midplane_temp} are box-like, low temperature (pressure) regions.  The odd shapes and ``x''-like markings seen in the Figure are artifacts of having a small number of cells within the feeding region of the stars. This happens because averaging over the cell can convert a small amount of kinetic energy into thermal energy.  Since the former is much larger than the latter in the wind-blown cavities, these small averaging errors lead to clearly visible features in the temperature. As the energy in these regions is dominated by the kinetic component by about 3 orders of magnitude, they have a negligible effect on the dynamics.

At late times (e.g., $t = 400$ kyr in the bottom right panel of Figure \ref{fig:midplane_density}), there is an accumulation of gas near the center of the simulation. This is, in fact, an accretion disk that forms from matter liberated from the CND, as we will show later. To see this disk more clearly, Figure \ref{fig:small_disk_times} shows a further zoom-in of the midplane density at an even smaller scale, $0.3 \times 0.3$ pc. This accretion disk is much denser and colder than the surrounding stellar wind material, with a temperature of $10^5$ K. This is hotter than the initial temperature of the disk but much cooler than the temperature of the surrounding shocked stellar wind material.
The process of inflow of high density matter to smaller radii takes several orbital periods at 1 pc - $T_{orb}(1 \mbox{ pc }) \sim 30$ kyr.

\subsection{Inflow and Outflow} \label{Results:Inflow and Outflow}

\begin{figure}
    \centering
    \includegraphics[width=0.95\textwidth]{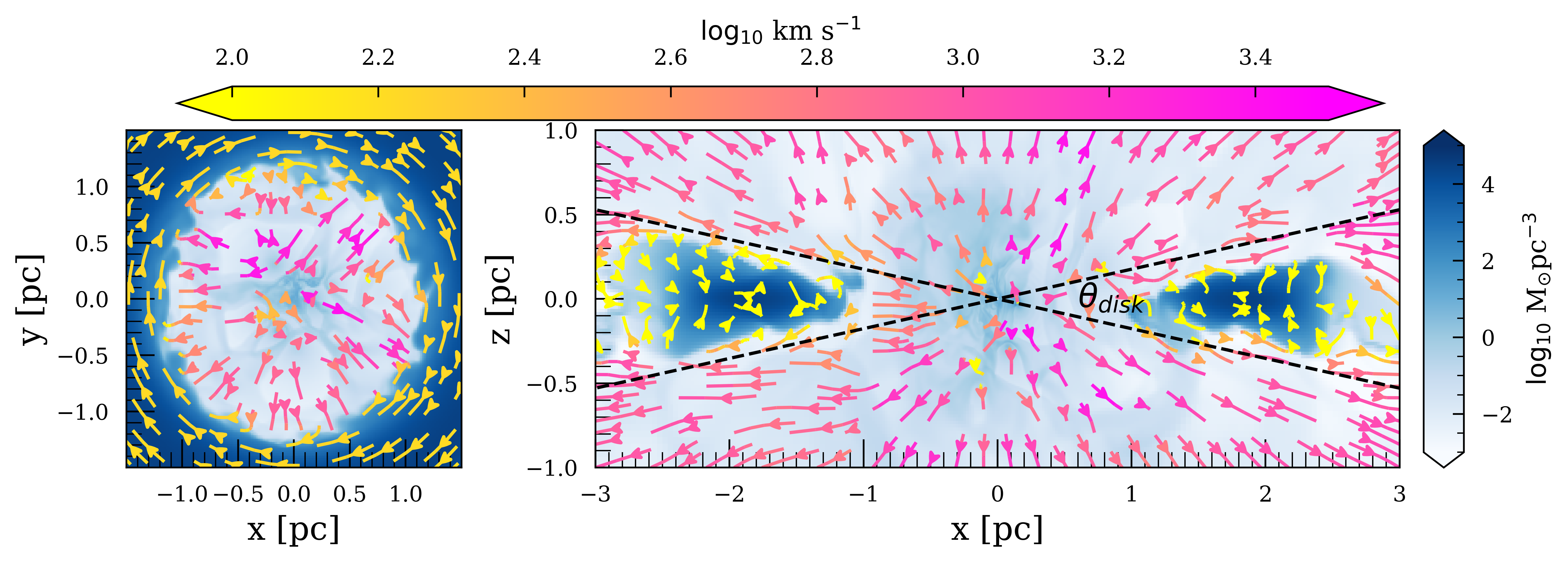}
    \caption{Slices of density at constant $z$ (left) and constant $y$ (right) overlaid with the velocity field of the fluid. The black dashed lines in the right panel ($x-z$ plane) represent the polar angle, $\theta_{disk} = 20^{\circ}$, that contains all of the disk.
    Note that the $x$ and the $z$ axes in the right panel are not to scale to more easily see flow structure. The combined flow of stellar wind material in the left panel ($x-y$ plane) has a radial velocity that meets with the inner edge of the disk, at which point there is a velocity discontinuity. }
    \label{fig:streamplot}
\end{figure}

We show the flow of gas in Figure \ref{fig:streamplot} which has the velocity field plotted over density slices in the $x-y$ (left) and $x-z$ (right) plane of the disc at $t=100$ kyrs. The angle subtended by the disk is $\theta_{disk} = 20^{\circ}$. The winds from the stars are a factor of $\sim 10$ faster than the orbital velocity of the disk. Most of the wind material does not intercept the CND and escapes outwards.

\begin{figure}
    \centering
    \includegraphics[width=0.95\textwidth]{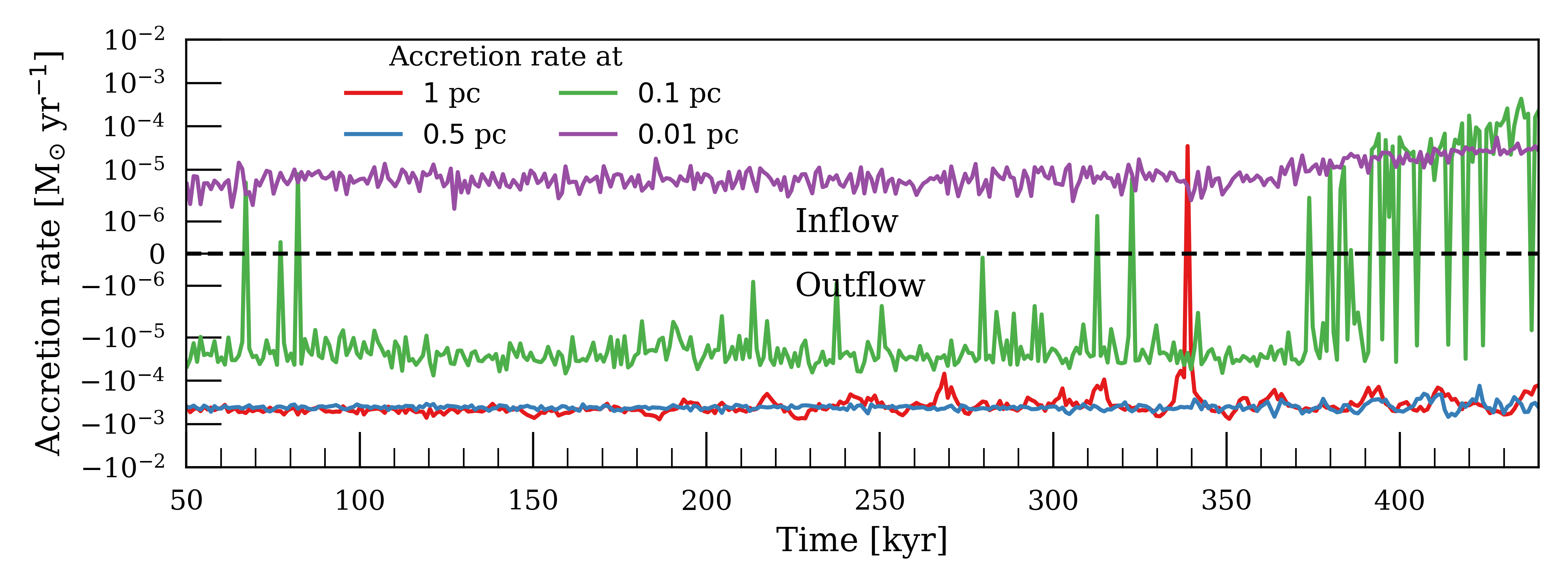}
    \caption{Net accretion rate as a function of time at different radii. The curves at $r=1$ (red), $0.5$ (blue) and $0.01$ pc (purple) remain fairly steady as most of the inflow/outflow comes from the gas provided by the stellar winds at all times. 
    However, at $r=0.1$ pc (green), there is a sharp transition from outflow to inflow at $\sim$ 390 kyr caused by the onset of accretion from the CND. 
    At the smallest radii (e.g., $r=0.01$ pc), this results in a factor of $\sim$ 2 increase in the accretion rate.}
    \label{fig:total_accretion_rate}
\end{figure}

We plot the net accretion rate on spheres of different radii as a function of time from $t = 50 $ to $440 $ kyrs in Figure \ref{fig:total_accretion_rate}, after allowing the inner edge of the CND to complete at least one orbit around the black hole. The accretion rate is computed over spherical shells located at $r = 1$, $0.5$, $0.1$  and $ 0.01$ pc. At larger radii ($r = 1$ and $r = 0.5$ pc), the flow of gas represents a steady outflow at a rate of $0.8$ M$_{\odot}$ kyr$^{-1}$ with the exception of a brief accretion event occurring at $340$ kyr in the $1$ pc curve. This outflow originates from the WR stellar winds, most of which have orbits that lie within $0.5$ pc. The winds of these stars are highly supersonic and mostly gravitationally unbound. Together the winds combine to form a global outflow that approaches a \citet{parker_wind} wind before colliding with the inner edge of the disk. The mass-loss from the stellar wind outflow is much greater than the inflow rate from the disk for all times at $r \gtrsim 1$ pc and therefore the net accretion rate at this radius is dominated by outflow. On top of this outflow, there is a small amplitude periodicity at the orbital period that is likely caused by accretion of the material from the CND.

At $r=0.1$ pc, however, inflow from the CND eventually grows larger in magnitude than outflow from the stellar winds after a certain transition time. This is seen in the corresponding curve in Figure \ref{fig:total_accretion_rate}, where the roughly constant $10^{-5}$ M$_{\odot}$ outflow at $0.1$ pc turns into inflow after $t \gtrsim 360$ kyrs.
Only a small fraction of the inflowing matter makes it to smaller radii $(r < 0.01 \text{pc})$, however; most of it accumulates and forms a small accretion disk. At yet smaller radii ($r=0.01$ pc), the flow is a steady inflow from the stellar winds.

\subsection{Inner Edge Instabilities} \label{Results:Inner Edge Instabilities}

\begin{figure}
    \centering
    \begin{tabular}{cc}
    {\includegraphics[width=0.45\columnwidth]{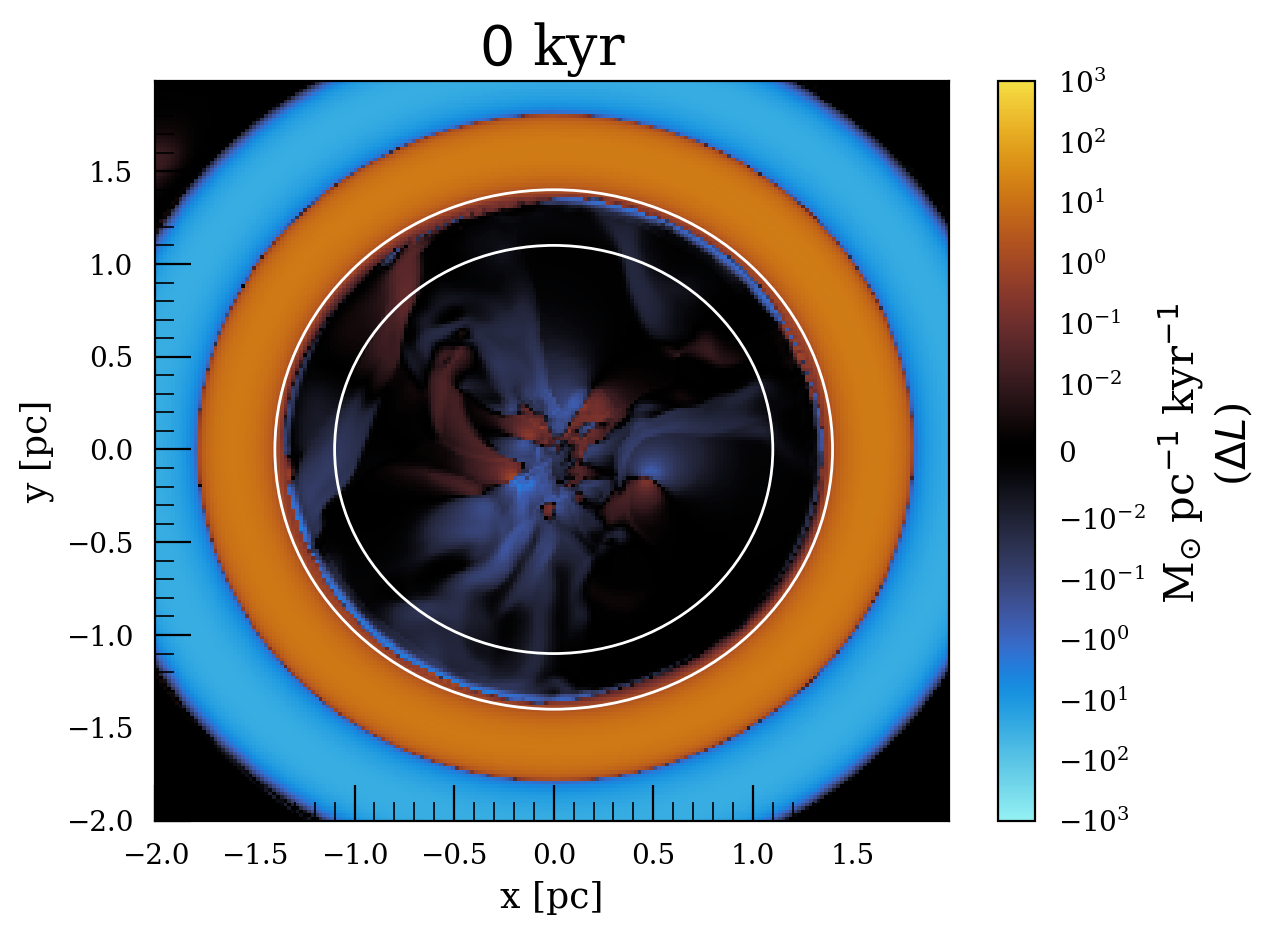}} &
    {\includegraphics[width=0.45\columnwidth]{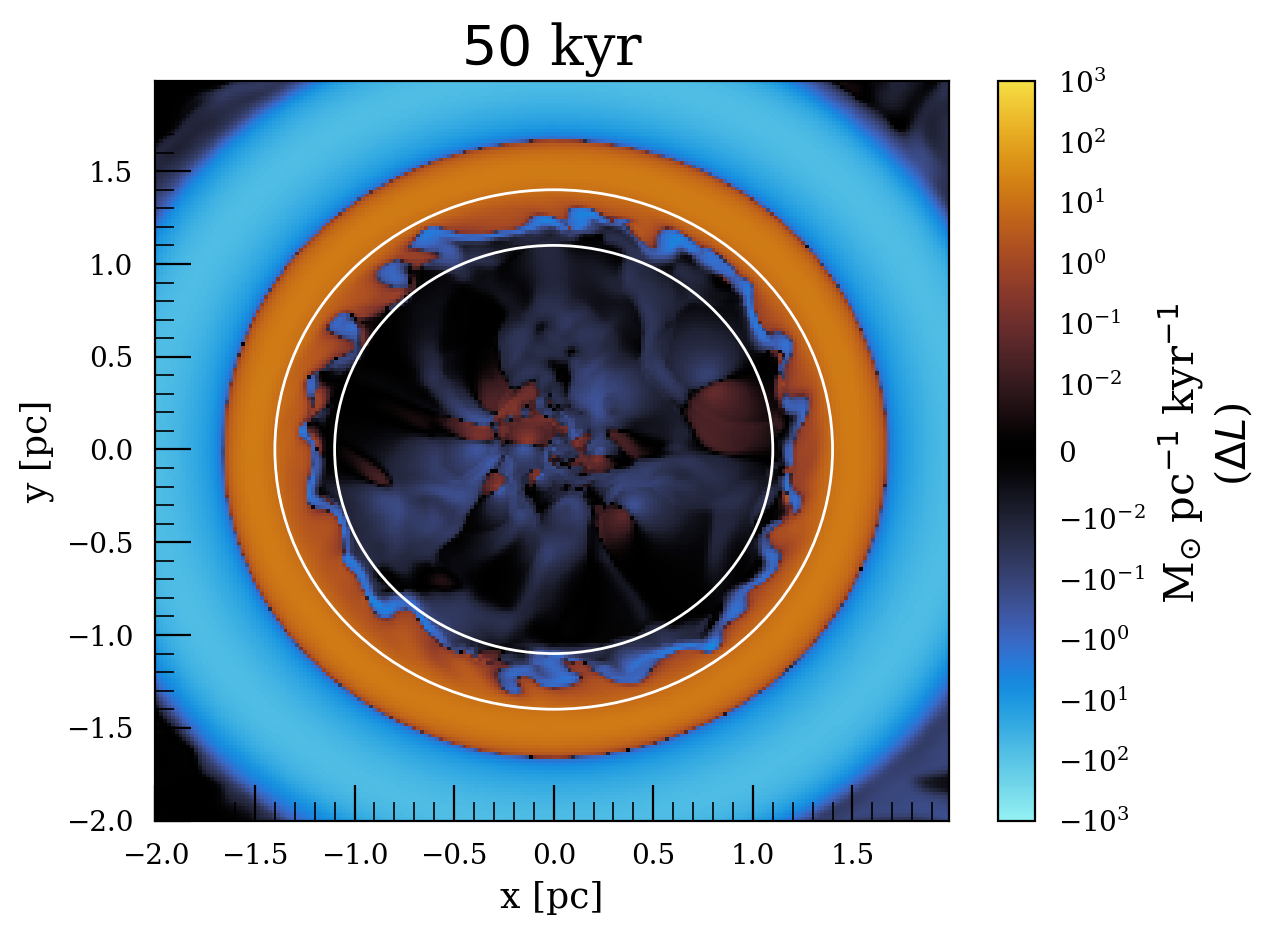}}
    \\
    {\includegraphics[width=0.45\columnwidth]{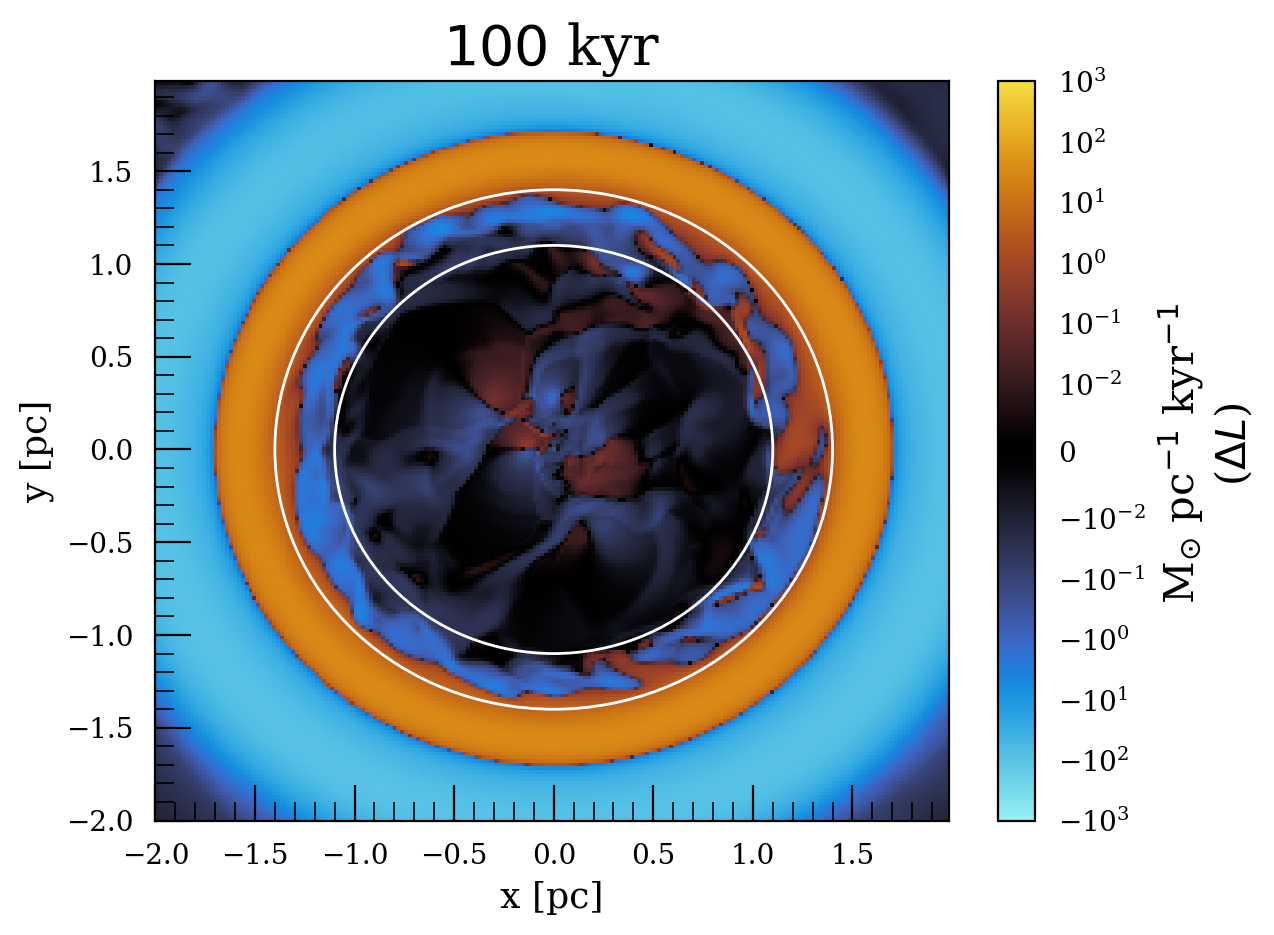}} &
    {\includegraphics[width=0.45\columnwidth]{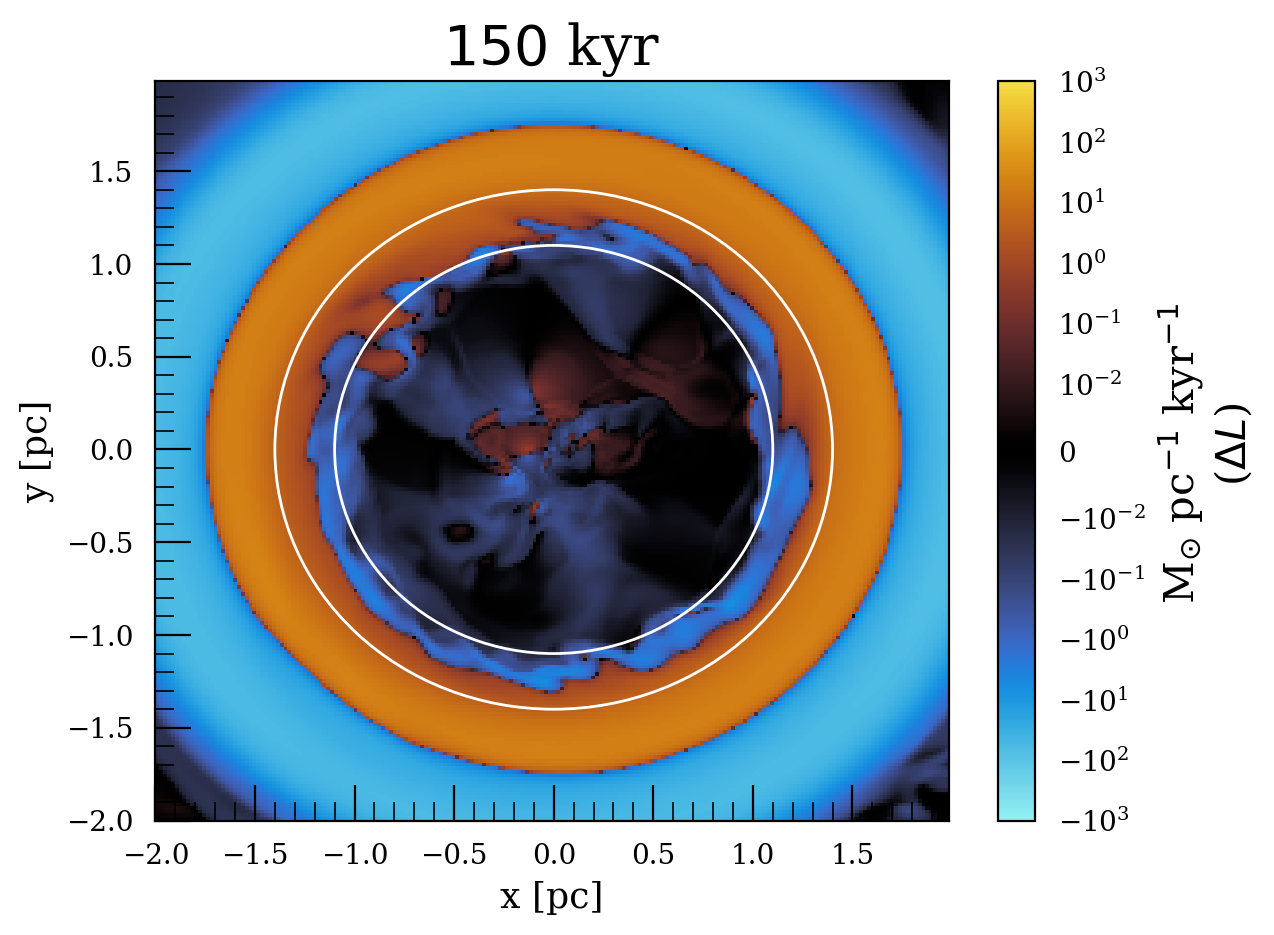}}
    \end{tabular}
    \caption{Difference in angular momentum of the gas in the simulation from gas in circular orbit with no external forces, $\Delta L$. Orange regions represent super-geodesic motion and the blue regions represent sub-geodesic motion. 
    The two white circles are drawn for static references at $r=1.4$ and $1.1$ pc respectively.
    Low angular momentum matter builds up on the inner edge of the disk as time passes, causing turbulent mixing and inwards migration.}
    \label{fig:angular_mom}
\end{figure}

The gas flow in the inner edge of the CND becomes turbulent as a result of the drag from the stellar winds and eventually starts accreting matter onto smaller scales. This can be seen from tracking the angular momentum of the gas.
Figure \ref{fig:angular_mom} plots the difference between the angular momentum of the gas in the simulation and the angular momentum of circular orbits at each radius at difference times. When this quantity is positive, the gas is rotating at a ``super-geodesic'' speed (i.e., absent other forces it will tend to spiral outwards), $\left(v^2/R > d\Phi/dR\right)$, and when it is negative the gas is rotating at a ``sub-geodesic'' speed (i.e., absent other forces it will tend to spiral inwards), $\left(v^2/R < d\Phi/dR\right)$.
The orange regions show gas that is rotating at a super-geodesic speed while the blue regions have a sub-geodesic speed. 
The two white circles are drawn for reference to track the movement of the inner edge of the disk.
Since the initial disk has non-zero thermal pressure support, the gas goes from super-geodesic (orange) to sub-geodesic (blue) motion at the pressure maximum (located at $R_k = 1.8$ pc in the top left panel). 
The subsequent panels show the low angular momentum material from the stellar winds getting deposited onto the inner edge of the disk. 
By $t=50$ kyr (the top right panel), ``valleys" of low angular momentum material have collected from the winds that penetrate the inner edge of the disk and grow to a length scale of several tenths of a pc before mixing with the high angular momentum gas due to differential rotation. 
As time proceeds, the inner edge of the disk migrates inwards as more and more low angular momentum material turbulently mixes into the disk.

\subsection{High and Low Temperature Flows} \label{Results:High and Low Temperature Flows}

\begin{figure}
    \begin{tabular}{cc}
    \includegraphics[width=0.7\textwidth]{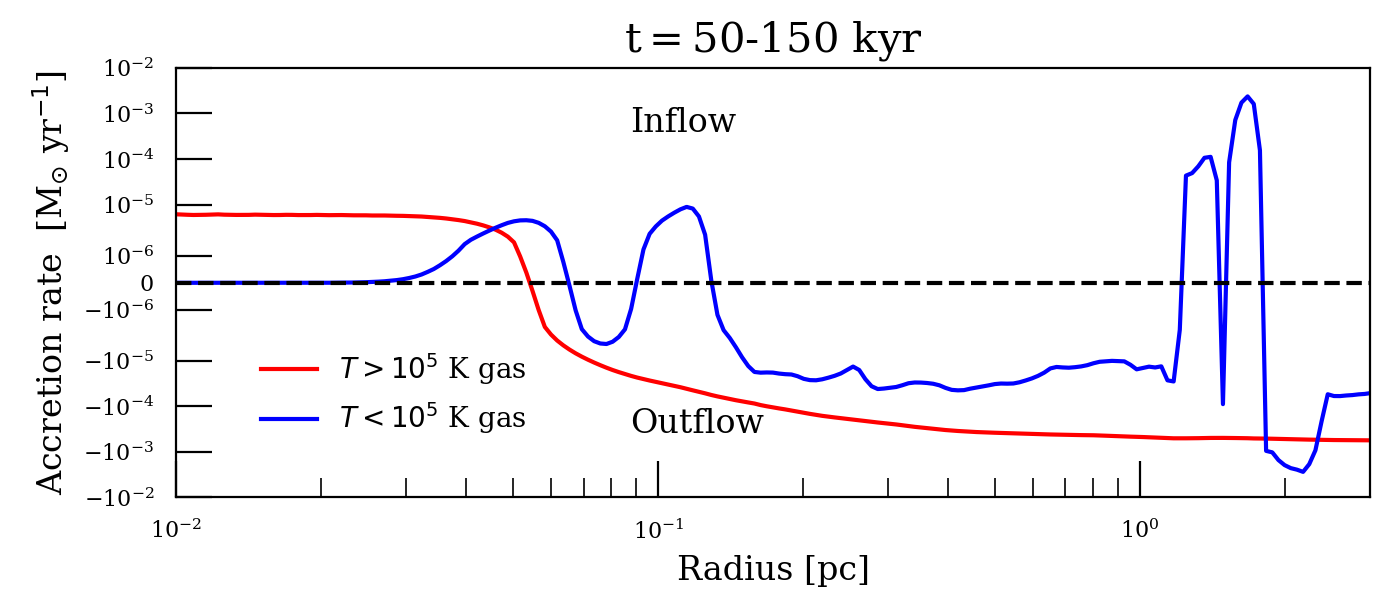} \\
    \includegraphics[width=0.7\textwidth]{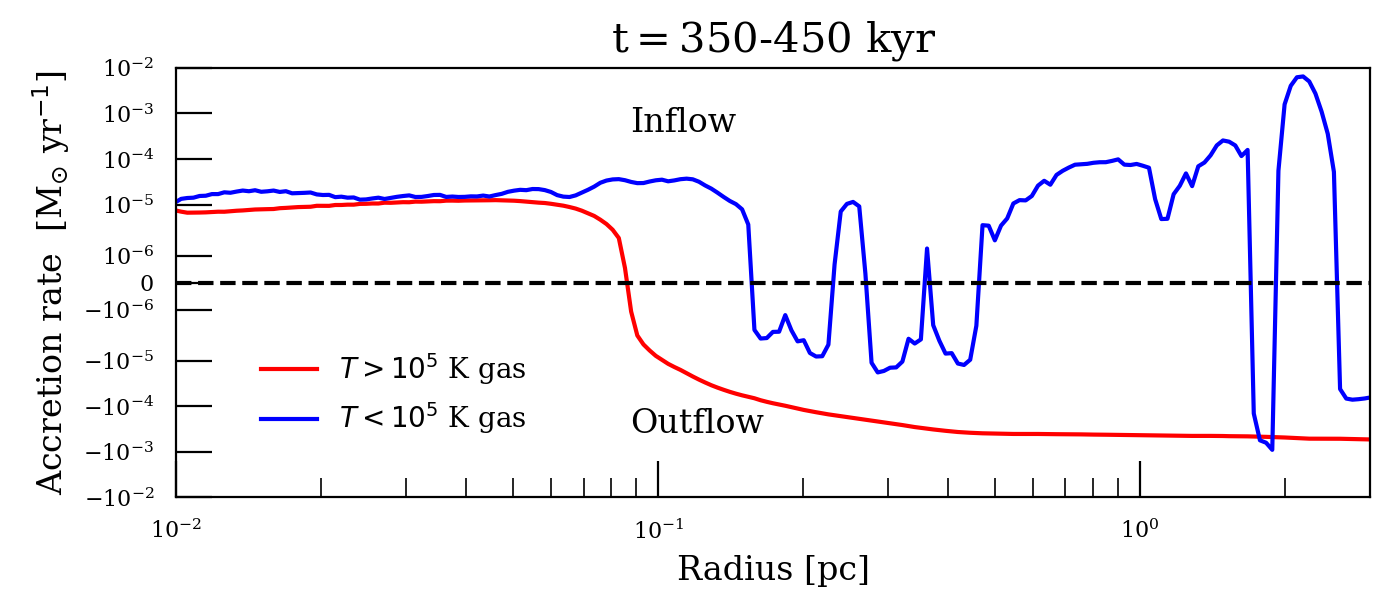}
    \end{tabular}
    \caption{The time-averaged accretion rates from high (red) and low (blue) temperature gas as a function of radius. The hotter gas ($T>10^5$ K) curve is much smoother and more constant in time than the colder gas because it is sourced by the WR stellar winds with orbital periods typically much smaller than the averaging time scale. 
    At $r=0.1$ pc there is a transition region from net inflow to outflow for the hot gas, representing roughly the location of the innermost WR stars.
     The cold gas ($T<10^5$ K) curve is much more variable in time and space on these timescales because it is sourced by the instabilities in the high density gas from the CND.
     At earlier times there is virtually no cold gas accretion to the smallest radii but at late times there is a steady inflow of cold gas for $r\lesssim 0.15$ pc roughly equal to the inflow of hot gas. }
    \label{fig:hot_cold_accretion}
\end{figure}

As mentioned in Section \ref{Results:Evolution of the Disk}, we are able to divide the gas in the simulation into two families based on its temperature for the purposes of analysis. In general, the hotter ($T > 10^5$ K) gas originates from the stellar winds and the ($T < 10^5$ K) gas originates from the disk. Here we focus on gas closer to the midplane by integrating the accretion rates over an angle of $\theta_{disk} = 20^{\circ}$ ($10^\circ$ on each side from the midplane) and a full $2{\rm \pi}$ in azimuth.
The time-averaged accretion rates for gas in both temperature ranges (integrated over $\theta_{disk}$) are shown in Figure \ref{fig:hot_cold_accretion}. 
The left and the right panels show time-averages from $t=50-100$ kyrs and $t=350-400$ kyrs respectively. 
In both the cases, the shape of the $T>10^5$ K gas curve is very similar. 
This is expected because most of the WR stellar winds lie within $0.5$ pc and so the time-average represents several orbital periods at that radius. 
Since the orbits of the stars and their mass loss rates stay the same for the entire simulation, there are not significant differences in the net wind flux. 
For this hotter gas, large radii outflow transitions to an inflow at $r \lesssim 0.05$ pc, which is within most of the stars' orbits. 
Within this radius the accretion rate is constant with radius at a value of $\sim 10^{-5}$ M$_{\odot}$ yr$^{-1}$ because there are no stars in that region so the flow reaches a steady state.
Conversely, virtually no low temperature gas is falling towards smaller scales at early times (the first panel). 
This changes after $300$ kyr once gas from the CND starts accreting. At this point, the accretion rates for $T>10^5$ K and $T<10^5$ K gas are comparable.

\begin{figure}
    \centering
    \includegraphics[width=0.7\textwidth]{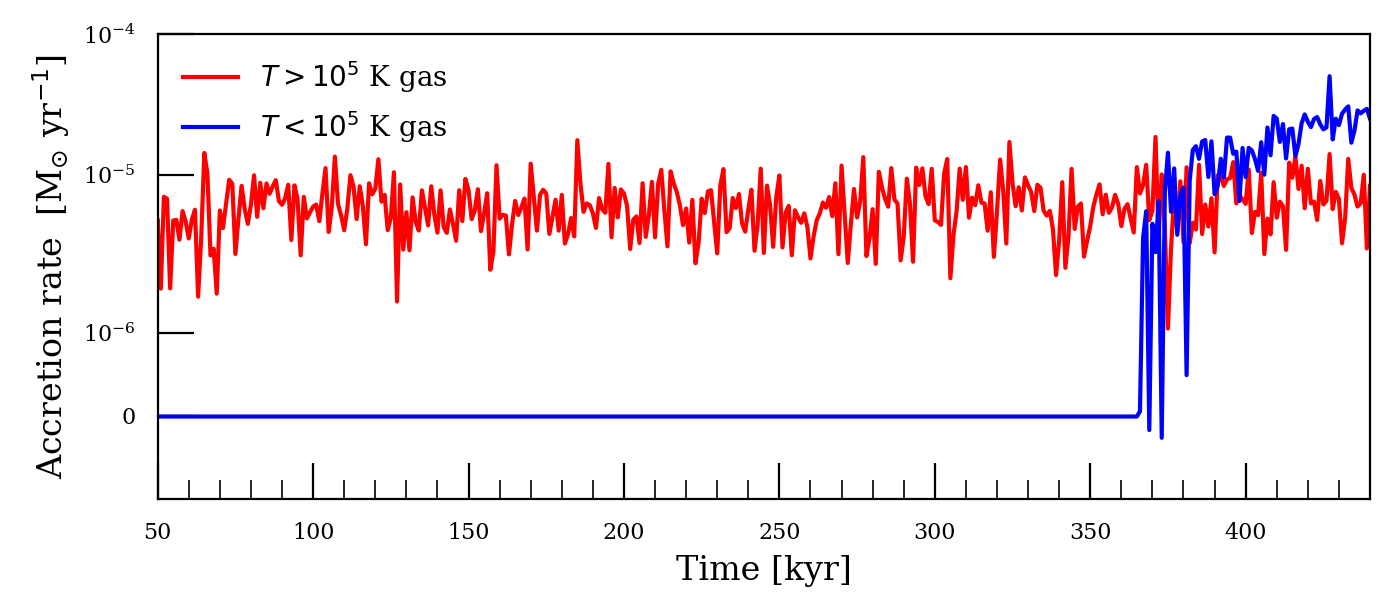}
    \caption{The high (red) and low (blue) temperature gas inflow rate, $\dot M_{\rm in}$ as a function of time at $r=0.01$ pc. The inflow from the stellar winds is relatively steady with time throughout the simulation. Cold gas only starts accreting at $\sim$ 350 kyr when gas from the CND reaches this radius. Cold gas inflow eventually grows even larger in magnitude than hot gas inflow at this radius.  }
    \label{fig:inflow_0.01_pc}
\end{figure}

To see exactly when this transition occurs, Figure \ref{fig:inflow_0.01_pc} shows the accretion rate at $r = 0.01$ pc as a function of time for the cold and hot gas separately. At this radius, the high temperature gas provided by the stellar winds has a roughly constant accretion rate of $10^{-5}$ M$_{\odot}$ kyr$^{-1}$.
Accretion from cold gas is negligible until $\sim$ 330 kyr, at which point it rapidly increases to a level comparable to and then even exceeding that from hot gas.

\subsection{Small Accretion Disk} \label{res:small_accretion_disk}

\begin{figure}
    \centering
    \includegraphics[width=0.95\columnwidth]{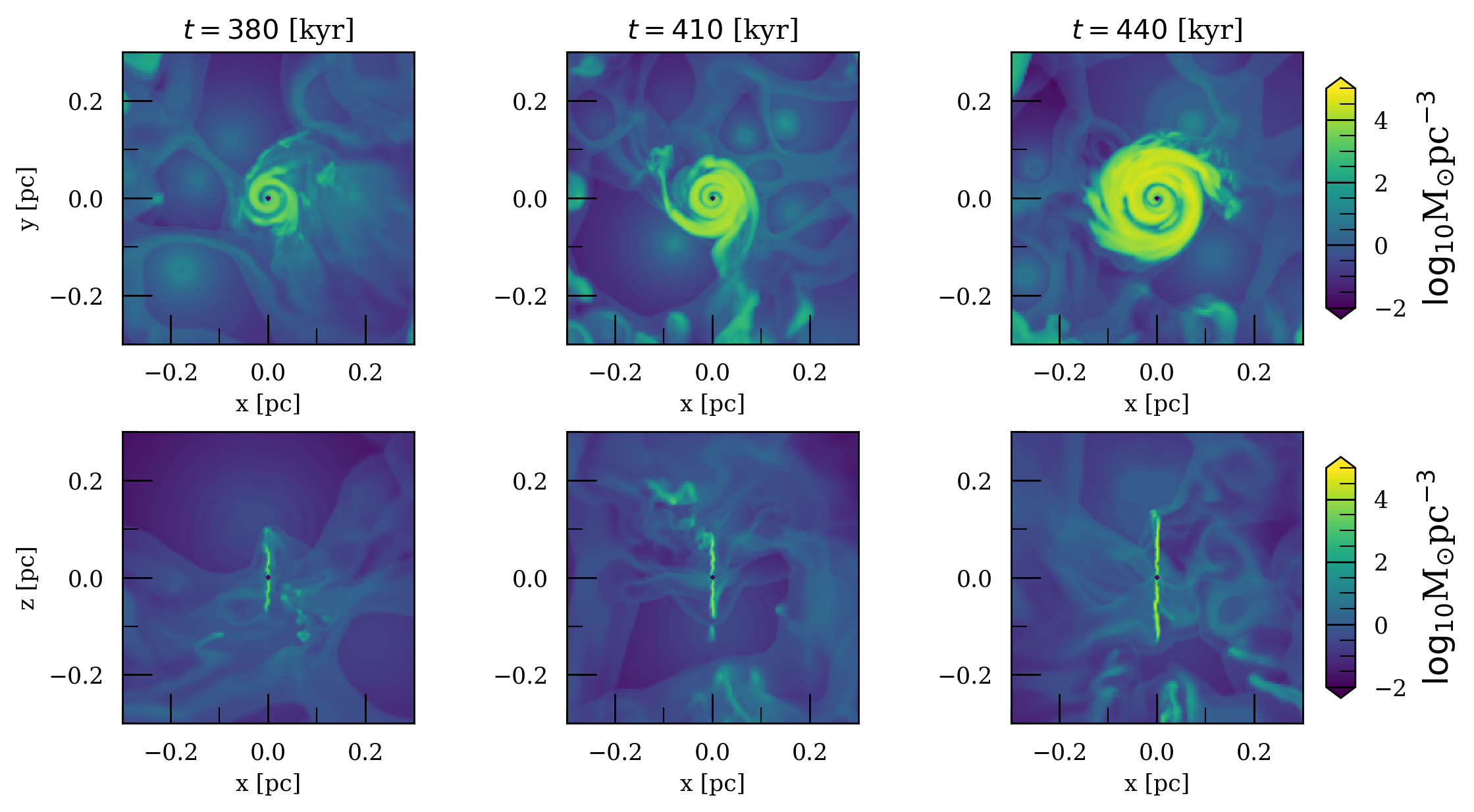}
    \caption{Density contours of the inner accretion disk at different times in the simulation. The disk is thin ($H/r \ll 1$) and comprised of cool gas ($T\sim 10^5$ K) accreted from the CND and is aligned with its angular momentum.  
    This smaller disk not only maintains its structure in the presence of the stellar winds but grows with time. 
    }
    \label{fig:small_disk_times}
\end{figure}

The infalling matter from the CND results in the formation of a small accretion disk around the central black hole. This disk consists of the colder matter from the CND and has a temperature of $T \sim 10^5$ K. We show the density slices of the accretion disk at different times in Figure \ref{fig:small_disk_times}. 
The angular momentum direction of this smaller disk is the same as the CND, that is, aligned with the $z$-direction.  We have tested that rotating the grid with respect to the CND that the smaller disk does not change this result; the smaller disk always aligns with the CND independently of the grid.
The disk is quite thin, with $H/r\ll 1$, but maintains its structure despite being surrounded by much hotter gas provided by the stellar winds.
In fact, the disk grows in size over time, nearly doubling in radial extent from $t = 380$ kyr to $t= 440$ kyr.  

The corresponding growth in mass of this smaller disk is plotted in Figure \ref{fig:small_disk_times} as a function of time.  We compute this quantity by summing all the mass in the inner $0.3$ pc$^3$ with densities $\rho > 10^3 \textrm{M}_{\odot} \text{ pc}^{-3}$.  Between $t=380$ kyr and $t = 440$ kyr (the end of the simulation) the mass grows from $\approx$ 0.1 $M_\odot$ to $2.3 \textrm{M}_{\odot}$.
At this point the disk mass still has not saturated and it is unclear whether it will continue to grow indefinitely.

\begin{figure}
    \centering
    \includegraphics[width=0.95\textwidth]{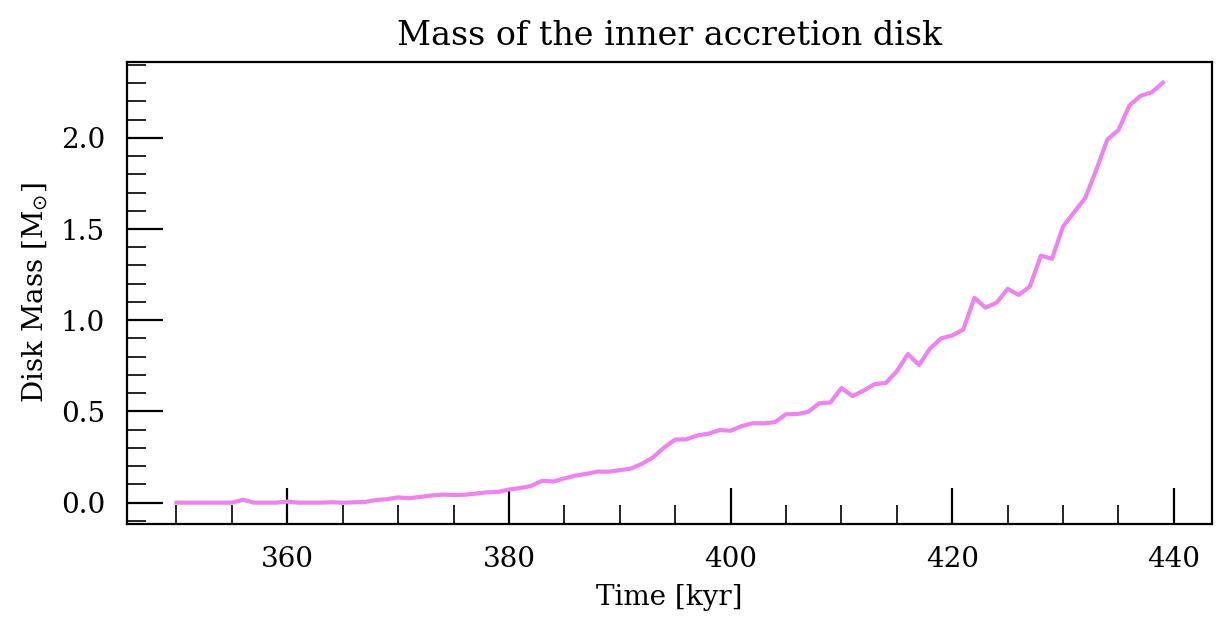}
    \caption{The mass of the smaller accretion disk as a function of time. The disk forms at around $350$ kyr and steadily grows as a result of CND accretion. The disk is not able to reach mass equilibrium by the end of the simulation.}
    \label{fig:inner_disk_mass}
\end{figure}

\section{Discussion} \label{Discussion}

\subsection{Angular Momentum Transport in the Disk}\label{Discussion:Angular Momentum Transport in the Disk}
Accretion from a disk in hydrodynamic equilibrium can only occur with some form of angular momentum transport. 
Here that transport is provided by the interaction between the low angular momentum stellar winds and the inner edge of the disk, as identified in \citet{10.1093/mnras/stw771}. Indeed, as seen in Figure \ref{fig:angular_mom}, low angular momentum matter builds up on the inner edge of the disk as it collects gas from the stellar winds. The orange regions show gas that is rotating at a super-geodesic speed $\left(v^2/R > d\Phi/dR\right)$ while the blue regions have $\left(v^2/R < d\Phi/dR\right)$. The wind from the stars produces a drag on the orbiting gas of the CND and the wind-disk interface becomes unstable to a Kelvin-Helmholtz like instability that occurs at the interface that causes mixing of the two types of gases. The inner edge also shows convective cycles of low and high angular momentum gas mixing. In Figure \ref{fig:angular_mom} this shows up as faster gas (orange) getting ahead of the slower gas (blue) as time progresses.  The mixing causes more gas to slow down and develop a lower specific angular momentum, ultimately forming a cogwheel-like pattern. 
In \citet{10.1093/mnras/stw771}, the authors computed the timescale of the inward migration of the inner edge of the disk based on the amount of low angular momentum wind material that gets mixed with the CND, obtaining a value of $2.2 \times 10^5$ kyrs. We observe similar timescales of $\sim 2-3 \times 10^5 $ yrs in our simulations. The aspect ratio of the disk is particularly important in determining the amount of low temperature gas that is present at smaller radii since the amount of the wind material intercepted is directly proportional to this quantity. 
A thicker disk will collect more wind material and will therefore accrete more material towards the center.

We ran another set of simulations with only the black hole gravitational potential (see Appendix \ref{Appendix_bh_only}). To summarize our findings, the presence of the NSC potential speeds up the instability growth and hence accretion from the CND starts at earlier times. The accretion rates at the innermost scales, however, turn out to be very similar with/without the NSC potential (Figure \ref{fig:accretion_rate_both}).
This is because (as mentioned above) the angular momentum transport (and hence the accretion) depends primarily on the amount of wind material the disk intercepts, which is similar in both cases.

\subsection{Comparison with Observations}\label{Discussion:Comparison with Observations}
\begin{figure}
    \centering
    \includegraphics[width=0.95\textwidth]{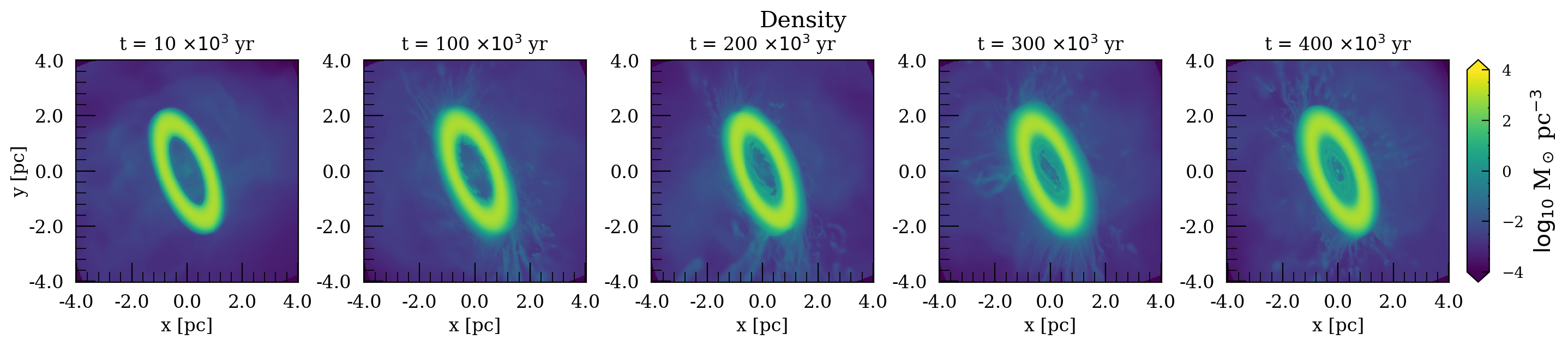}
    \includegraphics[width=0.95\textwidth]{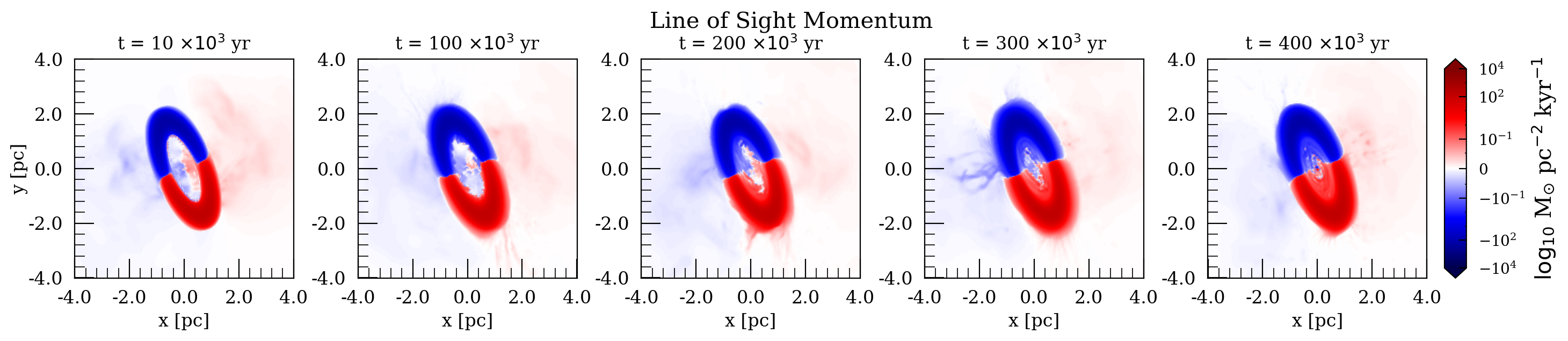}
    \includegraphics[width=0.95\textwidth]{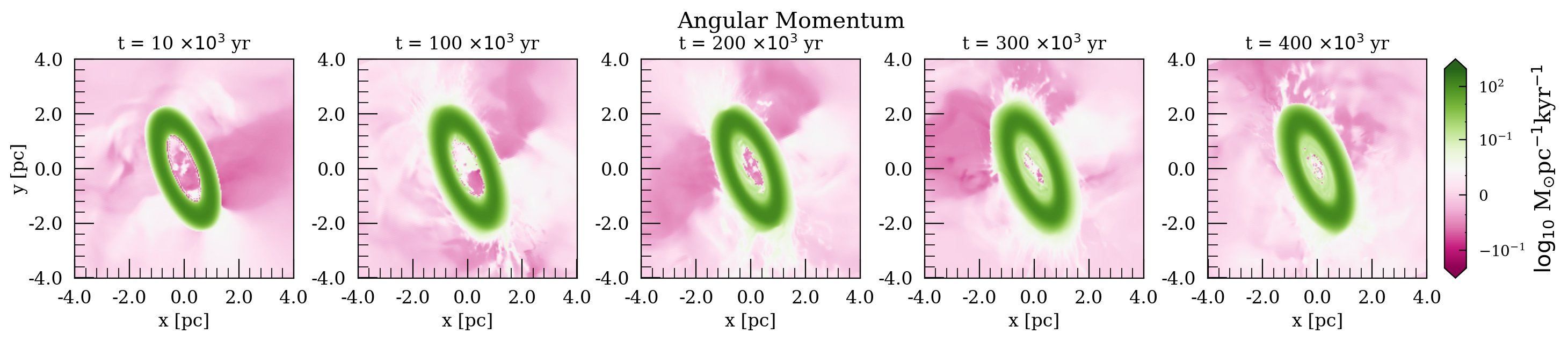}
    \caption{Slices of density (top), radial velocity (middle), and angular momentum (bottom) of gas in the plane of the sky. 
    In this frame the CND has a position angle of $22^{\circ}$ and an inclination of $66^{\circ}$.
    The cold gas in the CND dominates the momentum and the angular momentum of the simulation.}
    \label{fig:plane_of_sky}
\end{figure}

To more directly compare with observations, we plot density, linear momentum, and angular momentum averaged along the line of sight in Figure \ref{fig:plane_of_sky}. In our plots a positive angular momentum corresponds to a counter-clockwise rotation and a negative angular angular momentum corresponds to a clockwise rotation. 
In this new frame the projected major axis of the disk has a position angle of $22^{\circ}$ and inclination angle of $66^{\circ}$. 
The gas in the CND dominates the momentum and angular momentum shown in Figure \ref{fig:plane_of_sky} because of its high density. 
On the outer edge of the disk, streams of gas are visible that result from outward angular momentum transport through the disk, though these streams carry a very small fraction of the total angular momentum about the line of sight. 
On average, gas provided by the stellar winds has an angular momentum direction that is opposite to that of the CND, into the plane of the sky. 
This is because many of the Wolf-Rayet stars lie on a clockwise disk and the winds have a time-averaged angular momentum of the disk \citep{10.1093/mnras/sty1146}.

The small accretion disk that forms in our simulations (Figure \ref{fig:small_disk_times}) is qualitatively similar to the $r\sim 0.01$ pc disk observed in measurements of the H$30\alpha$ recombination line \citep{Murchikova2019}. 
Quantitatively, however, our disk is much larger in extent, reaching out to $\sim 0.2$ pc.
The amount of mass contained in the disk is an open question, since there is some tension with fact that it is not detected in Br$\gamma$ emission \citep{Ciurlo_2021}. 
One possibility for why our simulations may produce a larger disk than observed is that the magnetic pressure from toroidal magnetic fields in the Galactic Center could partially inhibit accretion at shorter timescales before the MRI instability from the smaller poloidal component kicks in \citep{10.1093/mnras/stw771}. Moreover, the presence of a toroidal magnetic field reduces the growth rate of the Kelvin-Helmholtz instability which has been shown to impede accretion \citep{10.1093/mnras/stw771}. Our purely hydrodynamics simulations would thus overpredict the accretion rate from the CND and the resulting size of the smaller disk.
Still, accretion from the CND represents a plausible explanation for how such a relatively cool disk could form.

\subsection{Limitations of the Model}\label{Discussion:Limitation of the Model}
The initial conditions of a disk in hydrodynamic equilibrium is a simplified model of the CND. Observations tell us that the CND and its surroundings have a much more complex morphology.

In reality, the CND is much colder than the one set up in the simulation, with temperatures ranging from $100-500$ K. As discussed in Section  \ref{Methods:Initializing the CND}, the larger temperature is necessary to maintain the observed aspect ratio of the disk. The high temperature in our simulations plays the role of the turbulent pressure in the CND.

While we expect that this higher temperature does not affect the dynamics of the simulation, our simulations do not correctly account for the thermodynamics of the CND. This is also because we do not track ionization rates nor account for different gas compositions in the CND vs. stellar wind-sourced gas and thus our cooling rates are not fully accurate.

Further complicating matters is the fact that The inner edge of the CND is ionized from stellar radiation

Our simulations do not include any magnetic fields. Observationally, the CND is known to have a well ordered, $\sim$ mG in strength magnetic field with a plasma $\beta \lesssim 1$ that may be relevant in the evolution of the CND \citep{Hsieh_2018}. 
\citet{10.1093/mnras/stw771} found that the presence of a purely toroidal magnetic field may actually slow down the migration of the inner edge of the disk, at least on relatively short timescales before the possible onset of MRI turbulence. 
Even so, the full geometry of the magnetic field in the CND also includes a weaker, poloidal component \citep{Hsieh_2018} and the resulting dynamics are plausibly quite sensitive to its precise configuration given the relatively high strength of the $\beta \lesssim 1$ field.

Another consideration is that The Galactic Center contains several massive stars that can go supernova in a $\sim$ $10^5$ yr timescale (approximately equal to the typical lifetime of the WR phase, \citealt{Crowther2007}). The simulations of \citet{Dinh_2021} showed that perturbations from supernovae can shear and fragment the CND, even causing the formation of self-gravitating clumps.

\section{Conclusions} \label{Conclusions}
We have presented hydrodynamic simulations of the Circumnuclear disk in the Galactic Center in the presence of stellar winds from Wolf-Rayet stars. The simulations include gravity from both Sgr A* and the Milky Way Nuclear Star Cluster and extend from $\sim$ 10 pc down to 0.002 pc, much further in than previous studies. 
The stellar winds produce a drag on the CND and cause its inner edge to become unstable and accrete towards smaller radii on a few $10^5$ year timescale due to turbulent angular momentum transport via the Kelvin-Helmholtz instability (Figures \ref{fig:midplane_density} and  \ref{fig:midplane_temp}).
The instability is caused by the interaction of the inner edge of the disk with the relatively low angular momentum stellar wind material (Figure \ref{fig:angular_mom}).

Accretion from the CND eventually leads to the formation of a small accretion disk after $\sim$ 350 kyr with a radial extent of $\sim 0.2$ pc (Figure \ref{fig:small_disk_times}. 
This time is comparable to the typical lifetime of the WR phase in massive stars and thus represents relatively late-stage evolution of the wind/CND system.
The disk is composed of $T\sim 10^5$ K gas that is much colder than the surrounding stellar wind material but much hotter than the CND. Once formed, the disk continually grows with time (Figure \ref{fig:inner_disk_mass}), and eventually the inflow rate of this cold gas becomes even larger than the outflow rate from the stellar winds (Figure \ref{fig:hot_cold_accretion}). 
Although this disk seen in our simulations is larger in size than the reported cool disk in the Galactic Center, accretion from the CND represents a plausible explanation for its existence, especially given the uncertainties in our model that do not contain magnetic fields, which could inhibit accretion \citep{10.1093/mnras/stw771} and potentially result in a smaller disk.

We have also argued that accretion from the CND could explain the existence of the Galactic Center Mini-Spiral.
After the onset of accretion from the CND, our simulations do display some spiral-like streams of gas falling in from the CND of various sizes and shapes (e.g., Figures \ref{fig:midplane_density} and \ref{fig:angular_mom}). 
Given that the Mini-Spiral may be a relatively transient structure (at least on $\gtrsim$ kyr timescales), it is possible that it may represent one particular instance/realization of this phenomena.  

We have made several simplifying assumptions in our model of the CND.
For instance, we treat it as a smooth disk whereas observations show that the CND is composed of several streamers and possibly clumps.
We also do not include magnetic fields. The magnetic field in the CND is mostly poloidal and expected to be dynamically important, the presence of which may initially slow down the growth of the K-H instability \citep{shivamoggi1981kelvin}.
We have discussed in detail how each of these assumptions could effect our results in Section \ref{Discussion:Limitation of the Model}.  
In any case, we expect our model to be a good first-order approximation to the stellar wind/CND Galactic Center system and represents perhaps the most realistic simulation to date of this region.

\section*{Acknowledgements} \label{Acknowledgements}
SS thanks Omer Blaes, Chris White and participants at the GCOI workshop held at UCLA for insightful discussions and comments. SS thanks Jeremy Schnittman and Sasha Philippov for supporting this study.
SMR was supported by the Gordon and Betty Moore
Foundation through Grant GBMF7392. 
This research was supported in part by the National
Science Foundation (NSF) under Grant No. NSF PHY-1748958.
LM work on this publication is supported in part by Black Hole Initiative at Harvard University which is funded by the Gordon and Betty Moore Foundation, and  also made possible through the support of a grant from the John Templeton Foundation. The opinions expressed in this publication are those of the author(s) and do not necessarily reflect the views of these Foundations.
LM acknowledges the support of William D. Loughlin and Corning Glass Foundation Memberships at the Institute for Advanced Study.

This work was made possible by
computing time granted by the IAS on the Helios cluster.

\software{\texttt{Athena++} \citep{Stone_2020,White_2016}}

\clearpage

\bibliography{bibliography}{}

\begin{thebibliography}{}
\expandafter\ifx\csname natexlab\endcsname\relax\def\natexlab#1{#1}\fi
\providecommand{\url}[1]{\href{#1}{#1}}
\providecommand{\dodoi}[1]{doi:~\href{http://doi.org/#1}{\nolinkurl{#1}}}
\providecommand{\doeprint}[1]{\href{http://ascl.net/#1}{\nolinkurl{http://ascl.net/#1}}}
\providecommand{\doarXiv}[1]{\href{https://arxiv.org/abs/#1}{\nolinkurl{https://arxiv.org/abs/#1}}}

\bibitem[{{Becklin} {et~al.}(1982){Becklin}, {Gatley}, \&
  {Werner}}]{1982ApJ...258..135B}
{Becklin}, E.~E., {Gatley}, I., \& {Werner}, M.~W. 1982, \apj, 258, 135,
  \dodoi{10.1086/160060}

\bibitem[{Blank {et~al.}(2016)Blank, Morris, Frank, Carroll-Nellenback, \&
  Duschl}]{10.1093/mnras/stw771}
Blank, M., Morris, M.~R., Frank, A., Carroll-Nellenback, J.~J., \& Duschl,
  W.~J. 2016, Monthly Notices of the Royal Astronomical Society, 459, 1721,
  \dodoi{10.1093/mnras/stw771}

\bibitem[{{Blank} {et~al.}(2016){Blank}, {Morris}, {Frank},
  {Carroll-Nellenback}, \& {Duschl}}]{2016MNRAS.459.1721B}
{Blank}, M., {Morris}, M.~R., {Frank}, A., {Carroll-Nellenback}, J.~J., \&
  {Duschl}, W.~J. 2016, \mnras, 459, 1721, \dodoi{10.1093/mnras/stw771}

\bibitem[{{Calder{\'o}n} {et~al.}(2020){Calder{\'o}n}, {Cuadra}, {Schartmann},
  {Burkert}, \& {Russell}}]{Calderon_2020ApJ...888L...2C}
{Calder{\'o}n}, D., {Cuadra}, J., {Schartmann}, M., {Burkert}, A., \&
  {Russell}, C. M.~P. 2020, \apjl, 888, L2, \dodoi{10.3847/2041-8213/ab5e81}

\bibitem[{{Chatzopoulos} {et~al.}(2015){Chatzopoulos}, {Fritz}, {Gerhard},
  {Gillessen}, {Wegg}, {Genzel}, \& {Pfuhl}}]{2015MNRAS.447..948C}
{Chatzopoulos}, S., {Fritz}, T.~K., {Gerhard}, O., {et~al.} 2015, \mnras, 447,
  948, \dodoi{10.1093/mnras/stu2452}

\bibitem[{Ciurlo {et~al.}(2021)Ciurlo, Morris, Campbell, Ghez, Do, \&
  Chu}]{Ciurlo_2021}
Ciurlo, A., Morris, M.~R., Campbell, R.~D., {et~al.} 2021, The Astrophysical
  Journal, 910, 143, \dodoi{10.3847/1538-4357/abe71a}

\bibitem[{{Ciurlo} {et~al.}(2016){Ciurlo}, {Paumard}, {Rouan}, \&
  {Cl{\'e}net}}]{2016A&A...594A.113C}
{Ciurlo}, A., {Paumard}, T., {Rouan}, D., \& {Cl{\'e}net}, Y. 2016, \aap, 594,
  A113, \dodoi{10.1051/0004-6361/201527173}

\bibitem[{{Coker} \& {Melia}(1997)}]{1997ApJ...488L.149C}
{Coker}, R.~F., \& {Melia}, F. 1997, \apjl, 488, L149, \dodoi{10.1086/310925}

\bibitem[{{Crowther}(2007)}]{Crowther2007}
{Crowther}, P.~A. 2007, \araa, 45, 177,
  \dodoi{10.1146/annurev.astro.45.051806.110615}

\bibitem[{Cuadra {et~al.}(2008)Cuadra, Nayakshin, \& Martins}]{Cuadra:2007ba}
Cuadra, J., Nayakshin, S., \& Martins, F. 2008, Mon. Not. Roy. Astron. Soc.,
  383, 458, \dodoi{10.1111/j.1365-2966.2007.12573.x}

\bibitem[{Dinh {et~al.}(2021)Dinh, Salas, Morris, \& Naoz}]{Dinh_2021}
Dinh, C.~K., Salas, J.~M., Morris, M.~R., \& Naoz, S. 2021, The Astrophysical
  Journal, 920, 79, \dodoi{10.3847/1538-4357/ac185b}

\bibitem[{{Do} {et~al.}(2019){Do}, {Hees}, {Ghez}, {Martinez}, {Chu}, {Jia},
  {Sakai}, {Lu}, {Gautam}, {O'Neil}, {Becklin}, {Morris}, {Matthews},
  {Nishiyama}, {Campbell}, {Chappell}, {Chen}, {Ciurlo}, {Dehghanfar},
  {Gallego-Cano}, {Kerzendorf}, {Lyke}, {Naoz}, {Saida}, {Sch{\"o}del},
  {Takahashi}, {Takamori}, {Witzel}, \& {Wizinowich}}]{Do2019}
{Do}, T., {Hees}, A., {Ghez}, A., {et~al.} 2019, Science, 365, 664,
  \dodoi{10.1126/science.aav8137}

\bibitem[{Do {et~al.}(2019)Do, Hees, Ghez, Martinez, Chu, Jia, Sakai, Lu,
  Gautam, O’Neil, Becklin, Morris, Matthews, Nishiyama, Campbell, Chappell,
  Chen, Ciurlo, Dehghanfar, Gallego-Cano, Kerzendorf, Lyke, Naoz, Saida,
  Schödel, Takahashi, Takamori, Witzel, \&
  Wizinowich}]{doi:10.1126/science.aav8137}
Do, T., Hees, A., Ghez, A., {et~al.} 2019, Science, 365, 664,
  \dodoi{10.1126/science.aav8137}

\bibitem[{{Eckart} \& {Genzel}(1996)}]{eckart_S_Stars}
{Eckart}, A., \& {Genzel}, R. 1996, \nat, 383, 415, \dodoi{10.1038/383415a0}

\bibitem[{{Eckart} \& {Genzel}(1997)}]{Eckart_Genzel_1997MNRAS.284..576E}
---. 1997, \mnras, 284, 576, \dodoi{10.1093/mnras/284.3.576}

\bibitem[{{Einfeldt}(1988)}]{Einfeldt1988}
{Einfeldt}, B. 1988, SIAM Journal on Numerical Analysis, 25, 294,
  \dodoi{10.1137/0725021}

\bibitem[{Feldmeier {et~al.}(2014)Feldmeier, Neumayer, Seth, Schödel,
  Lützgendorf, de~Zeeuw, Kissler-Patig, Nishiyama, \&
  Walcher}]{Feldmeier_2014}
Feldmeier, A., Neumayer, N., Seth, A., {et~al.} 2014, Astronomy \&
  Astrophysics, 570, A2, \dodoi{10.1051/0004-6361/201423777}

\bibitem[{{Feldmeier-Krause} {et~al.}(2017){Feldmeier-Krause}, {Zhu},
  {Neumayer}, {van de Ven}, {de Zeeuw}, \& {Sch{\"o}del}}]{2017MNRAS.466.4040F}
{Feldmeier-Krause}, A., {Zhu}, L., {Neumayer}, N., {et~al.} 2017, \mnras, 466,
  4040, \dodoi{10.1093/mnras/stw3377}

\bibitem[{{Fishbone} \& {Moncrief}(1976)}]{GRMHD_torii_1}
{Fishbone}, L.~G., \& {Moncrief}, V. 1976, \apj, 207, 962,
  \dodoi{10.1086/154565}

\bibitem[{{Gallego-Cano} {et~al.}(2020){Gallego-Cano}, {Sch{\"o}del},
  {Nogueras-Lara}, {Dong}, {Shahzamanian}, {Fritz}, {Gallego-Calvente}, \&
  {Neumayer}}]{MWNSC_1}
{Gallego-Cano}, E., {Sch{\"o}del}, R., {Nogueras-Lara}, F., {et~al.} 2020,
  \aap, 634, A71, \dodoi{10.1051/0004-6361/201935303}

\bibitem[{{Genzel}(1989)}]{1989IAUS..136..393G}
{Genzel}, R. 1989, in The Center of the Galaxy, ed. M.~{Morris}, Vol. 136, 393

\bibitem[{{Genzel} {et~al.}(2010){Genzel}, {Eisenhauer}, \&
  {Gillessen}}]{Genzel10}
{Genzel}, R., {Eisenhauer}, F., \& {Gillessen}, S. 2010, Reviews of Modern
  Physics, 82, 3121, \dodoi{10.1103/RevModPhys.82.3121}

\bibitem[{{Genzel} {et~al.}(1994){Genzel}, {Hollenbach}, \&
  {Townes}}]{Genzel_1994RPPh...57..417G}
{Genzel}, R., {Hollenbach}, D., \& {Townes}, C.~H. 1994, Reports on Progress in
  Physics, 57, 417, \dodoi{10.1088/0034-4885/57/5/001}

\bibitem[{{Ghez} {et~al.}(1998{\natexlab{a}}){Ghez}, {Klein}, {Morris}, \&
  {Becklin}}]{Ghez_1998ApJ...509..678G}
{Ghez}, A.~M., {Klein}, B.~L., {Morris}, M., \& {Becklin}, E.~E.
  1998{\natexlab{a}}, \apj, 509, 678, \dodoi{10.1086/306528}

\bibitem[{{Ghez} {et~al.}(1998{\natexlab{b}}){Ghez}, {Klein}, {Morris}, \&
  {Becklin}}]{ghez_s_stars}
---. 1998{\natexlab{b}}, \apj, 509, 678, \dodoi{10.1086/306528}

\bibitem[{{Ghez} {et~al.}(2005){Ghez}, {Salim}, {Hornstein}, {Tanner}, {Lu},
  {Morris}, {Becklin}, \& {Duch{\^e}ne}}]{2005ApJ...620..744G}
{Ghez}, A.~M., {Salim}, S., {Hornstein}, S.~D., {et~al.} 2005, \apj, 620, 744,
  \dodoi{10.1086/427175}

\bibitem[{{Gravity Collaboration} {et~al.}(2019){Gravity Collaboration},
  {Abuter}, {Amorim}, {Baub{\"o}ck}, {Berger}, {Bonnet}, {Brandner},
  {Cl{\'e}net}, {Coud{\'e} Du Foresto}, {de Zeeuw}, {Dexter}, {Duvert},
  {Eckart}, {Eisenhauer}, {F{\"o}rster Schreiber}, {Garcia}, {Gao}, {Gendron},
  {Genzel}, {Gerhard}, {Gillessen}, {Habibi}, {Haubois}, {Henning}, {Hippler},
  {Horrobin}, {Jim{\'e}nez-Rosales}, {Jocou}, {Kervella}, {Lacour},
  {Lapeyr{\`e}re}, {Le Bouquin}, {L{\'e}na}, {Ott}, {Paumard}, {Perraut},
  {Perrin}, {Pfuhl}, {Rabien}, {Rodriguez Coira}, {Rousset}, {Scheithauer},
  {Sternberg}, {Straub}, {Straubmeier}, {Sturm}, {Tacconi}, {Vincent}, {von
  Fellenberg}, {Waisberg}, {Widmann}, {Wieprecht}, {Wiezorrek}, {Woillez}, \&
  {Yazici}}]{Gravity2019}
{Gravity Collaboration}, {Abuter}, R., {Amorim}, A., {et~al.} 2019, \aap, 625,
  L10, \dodoi{10.1051/0004-6361/201935656}

\bibitem[{Hsieh {et~al.}(2018)Hsieh, Koch, Kim, Ho, Tang, \& Wang}]{Hsieh_2018}
Hsieh, P.-Y., Koch, P.~M., Kim, W.-T., {et~al.} 2018, The Astrophysical
  Journal, 862, 150, \dodoi{10.3847/1538-4357/aacb27}

\bibitem[{Hsieh {et~al.}(2021)Hsieh, Koch, Kim, Mart{\'{\i} }n, Yen, Carpenter,
  Harada, Turner, Ho, Tang, \& Beck}]{Hsieh_2021}
---. 2021, The Astrophysical Journal, 913, 94, \dodoi{10.3847/1538-4357/abf4cd}

\bibitem[{Jackson {et~al.}(1993)Jackson, Geis, Genzel, Harris, Madden,
  Poglitsch, Stacey, \& Townes}]{jackson1993neutral}
Jackson, J., Geis, N., Genzel, R., {et~al.} 1993, The Astrophysical Journal,
  402, 173

\bibitem[{{James} {et~al.}(2021){James}, {Viti}, {Yusef-Zadeh}, {Royster}, \&
  {Wardle}}]{James2021}
{James}, T.~A., {Viti}, S., {Yusef-Zadeh}, F., {Royster}, M., \& {Wardle}, M.
  2021, \apj, 916, 69, \dodoi{10.3847/1538-4357/abfd99}

\bibitem[{{Kim} \& {Ostriker}(2017)}]{2017ApJ...846..133K}
{Kim}, C.-G., \& {Ostriker}, E.~C. 2017, \apj, 846, 133,
  \dodoi{10.3847/1538-4357/aa8599}

\bibitem[{Koyama \& Inutsuka(2001)}]{Koyama_2001}
Koyama, H., \& Inutsuka. 2001, The Astrophysical Journal, 564, L97,
  \dodoi{10.1086/338978}

\bibitem[{{Krabbe} {et~al.}(1995){Krabbe}, {Genzel}, \&
  {Eckart}}]{Krabbe__1995}
{Krabbe}, A., {Genzel}, R., \& {Eckart}, A. 1995, \apjl, 447,
  \dodoi{10.1086/309579}

\bibitem[{{Kunneriath} {et~al.}(2012){Kunneriath}, {Eckart}, {Vogel}, {Teuben},
  {Mu{\v{z}}i{\'c}}, {Sch{\"o}del}, {Garc{\'\i}a-Mar{\'\i}n}, {Moultaka},
  {Staguhn}, {Straubmeier}, {Zensus}, {Valencia-S.}, \&
  {Karas}}]{2012A&A...538A.127K_mini_spiral}
{Kunneriath}, D., {Eckart}, A., {Vogel}, S.~N., {et~al.} 2012, \aap, 538, A127,
  \dodoi{10.1051/0004-6361/201117676}

\bibitem[{Lacy {et~al.}(1991)Lacy, Achtermann, \&
  Serabyn}]{lacy1991galactic_mini_spiral}
Lacy, J., Achtermann, J., \& Serabyn, E. 1991, The Astrophysical Journal, 380,
  L71

\bibitem[{{Levin} \&
  {Beloborodov}(2003)}]{Levin_Beloborodov_2003ApJ...590L..33L}
{Levin}, Y., \& {Beloborodov}, A.~M. 2003, \apjl, 590, L33,
  \dodoi{10.1086/376675}

\bibitem[{{Lo} \& {Claussen}(1983)}]{1983Natur.306..647L}
{Lo}, K.~Y., \& {Claussen}, M.~J. 1983, \nat, 306, 647,
  \dodoi{10.1038/306647a0}

\bibitem[{{Lu} {et~al.}(2013){Lu}, {Do}, {Ghez}, {Morris}, {Yelda}, \&
  {Matthews}}]{2013ApJ...764..155L}
{Lu}, J.~R., {Do}, T., {Ghez}, A.~M., {et~al.} 2013, \apj, 764, 155,
  \dodoi{10.1088/0004-637X/764/2/155}

\bibitem[{{Lu} {et~al.}(2009){Lu}, {Ghez}, {Hornstein}, {Morris}, {Becklin}, \&
  {Matthews}}]{lu_stars}
{Lu}, J.~R., {Ghez}, A.~M., {Hornstein}, S.~D., {et~al.} 2009, \apj, 690, 1463,
  \dodoi{10.1088/0004-637X/690/2/1463}

\bibitem[{{Martins} {et~al.}(2007){Martins}, {Genzel}, {Hillier}, {Eisenhauer},
  {Paumard}, {Gillessen}, {Ott}, \& {Trippe}}]{Martins_stars}
{Martins}, F., {Genzel}, R., {Hillier}, D.~J., {et~al.} 2007, \aap, 468, 233,
  \dodoi{10.1051/0004-6361:20066688}

\bibitem[{{Meynet} \& {Maeder}(2005)}]{WR_lifetime}
{Meynet}, G., \& {Maeder}, A. 2005, \aap, 429, 581,
  \dodoi{10.1051/0004-6361:20047106}

\bibitem[{{Morris} \& {Serabyn}(1996)}]{Morris_GC_review}
{Morris}, M., \& {Serabyn}, E. 1996, \araa, 34, 645,
  \dodoi{10.1146/annurev.astro.34.1.645}

\bibitem[{Murchikova {et~al.}(2019)Murchikova, Phinney, Pancoast, \&
  Blandford}]{Murchikova2019}
Murchikova, E.~M., Phinney, E.~S., Pancoast, A., \& Blandford, R.~D. 2019,
  Nature, 570, 83, \dodoi{10.1038/s41586-019-1242-z}

\bibitem[{{Naoz} {et~al.}(2018){Naoz}, {Ghez}, {Hees}, {Do}, {Witzel}, \&
  {Lu}}]{2018ApJ...853L..24N}
{Naoz}, S., {Ghez}, A.~M., {Hees}, A., {et~al.} 2018, \apjl, 853, L24,
  \dodoi{10.3847/2041-8213/aaa6bf}

\bibitem[{{Parker}(1965)}]{parker_wind}
{Parker}, E.~N. 1965, \ssr, 4, 666, \dodoi{10.1007/BF00216273}

\bibitem[{{Paumard} {et~al.}(2006){Paumard}, {Genzel}, {Martins}, {Nayakshin},
  {Beloborodov}, {Levin}, {Trippe}, {Eisenhauer}, {Ott}, {Gillessen}, {Abuter},
  {Cuadra}, {Alexander}, \& {Sternberg}}]{paumard_massive_stars}
{Paumard}, T., {Genzel}, R., {Martins}, F., {et~al.} 2006, \apj, 643, 1011,
  \dodoi{10.1086/503273}

\bibitem[{Penna {et~al.}(2013)Penna, Kulkarni, \& Narayan}]{GRMHD_torii_2}
Penna, R.~F., Kulkarni, A., \& Narayan, R. 2013, Astronomy {\&}amp
  Astrophysics, 559, A116, \dodoi{10.1051/0004-6361/201219666}

\bibitem[{Ressler {et~al.}(2018)Ressler, Quataert, \&
  Stone}]{10.1093/mnras/sty1146}
Ressler, S.~M., Quataert, E., \& Stone, J.~M. 2018, Monthly Notices of the
  Royal Astronomical Society, 478, 3544, \dodoi{10.1093/mnras/sty1146}

\bibitem[{{Ressler} {et~al.}(2020){Ressler}, {Quataert}, \&
  {Stone}}]{2020MNRAS.492.3272R}
{Ressler}, S.~M., {Quataert}, E., \& {Stone}, J.~M. 2020, \mnras, 492, 3272,
  \dodoi{10.1093/mnras/stz3605}

\bibitem[{Ressler {et~al.}(2020)Ressler, White, Quataert, \&
  Stone}]{Ressler_2020}
Ressler, S.~M., White, C.~J., Quataert, E., \& Stone, J.~M. 2020, The
  Astrophysical Journal, 896, L6, \dodoi{10.3847/2041-8213/ab9532}

\bibitem[{{Sch{\"o}del} {et~al.}(2014){Sch{\"o}del}, {Feldmeier}, {Neumayer},
  {Meyer}, \& {Yelda}}]{MWNSC_2}
{Sch{\"o}del}, R., {Feldmeier}, A., {Neumayer}, N., {Meyer}, L., \& {Yelda}, S.
  2014, Classical and Quantum Gravity, 31, 244007,
  \dodoi{10.1088/0264-9381/31/24/244007}

\bibitem[{Shivamoggi(1981)}]{shivamoggi1981kelvin}
Shivamoggi, B.~K. 1981, Applied Scientific Research, 37, 291

\bibitem[{Stone {et~al.}(2020)Stone, Tomida, White, \& Felker}]{Stone_2020}
Stone, J.~M., Tomida, K., White, C.~J., \& Felker, K.~G. 2020, The
  Astrophysical Journal Supplement Series, 249, 4,
  \dodoi{10.3847/1538-4365/ab929b}

\bibitem[{Townsend(2009)}]{Townsend_2009}
Townsend, R. H.~D. 2009, The Astrophysical Journal Supplement Series, 181, 391,
  \dodoi{10.1088/0067-0049/181/2/391}

\bibitem[{Tsuboi {et~al.}(2017)Tsuboi, Kitamura, Uehara, Miyawaki, Tsutsumi,
  Miyazaki, \& Miyoshi}]{Tsuboi_2017_MS_orbits}
Tsuboi, M., Kitamura, Y., Uehara, K., {et~al.} 2017, The Astrophysical Journal,
  842, 94, \dodoi{10.3847/1538-4357/aa74e3}

\bibitem[{{Tsuboi} {et~al.}(2017){Tsuboi}, {Kitamura}, {Uehara}, {Miyawaki},
  {Tsutsumi}, {Miyazaki}, \& {Miyoshi}}]{Tsuboi2017}
{Tsuboi}, M., {Kitamura}, Y., {Uehara}, K., {et~al.} 2017, \apj, 842, 94,
  \dodoi{10.3847/1538-4357/aa74e3}

\bibitem[{{Vollmer} \& {Duschl}(2001)}]{Clumoy_disk_1_2001A&A...367...72V}
{Vollmer}, B., \& {Duschl}, W.~J. 2001, \aap, 367, 72,
  \dodoi{10.1051/0004-6361:20000425}

\bibitem[{{Vollmer} \& {Duschl}(2002)}]{Clumpy_disk_2_2002A&A...388..128V}
---. 2002, \aap, 388, 128, \dodoi{10.1051/0004-6361:20020422}

\bibitem[{White {et~al.}(2016)White, Stone, \& Gammie}]{White_2016}
White, C.~J., Stone, J.~M., \& Gammie, C.~F. 2016, ApJS, 225, 22,
  \dodoi{10.3847/0067-0049/225/2/22}

\bibitem[{{Wright} {et~al.}(2001){Wright}, {Coil}, {McGary}, {Ho}, \&
  {Harris}}]{2001ApJ...551..254W}
{Wright}, M. C.~H., {Coil}, A.~L., {McGary}, R.~S., {Ho}, P. T.~P., \&
  {Harris}, A.~I. 2001, \apj, 551, 254, \dodoi{10.1086/320089}

\bibitem[{{Yelda} {et~al.}(2014){Yelda}, {Ghez}, {Lu}, {Do}, {Meyer}, {Morris},
  \& {Matthews}}]{2014ApJ...783..131Y}
{Yelda}, S., {Ghez}, A.~M., {Lu}, J.~R., {et~al.} 2014, \apj, 783, 131,
  \dodoi{10.1088/0004-637X/783/2/131}

\bibitem[{{Zhao} {et~al.}(2009){Zhao}, {Morris}, {Goss}, \& {An}}]{Zhao2009}
{Zhao}, J.-H., {Morris}, M.~R., {Goss}, W.~M., \& {An}, T. 2009, \apj, 699,
  186, \dodoi{10.1088/0004-637X/699/1/186}

\end{thebibliography}
\bibliographystyle{aasjournal}
\appendix
\section{Effect of the Galactic Potential} \label{Appendix_bh_only}
\begin{figure}
    \centering
    \includegraphics[width=0.95\textwidth]{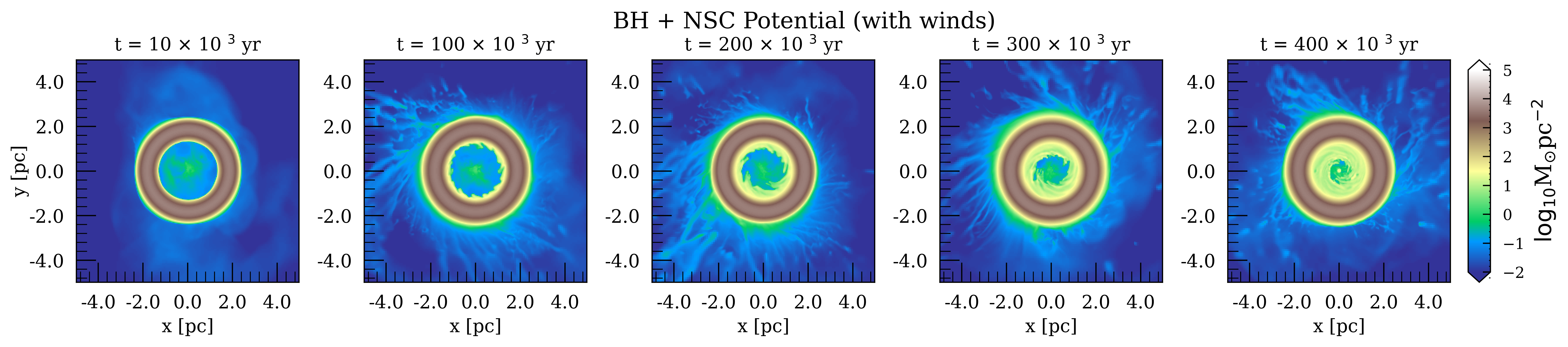}
    \includegraphics[width=0.95\textwidth]{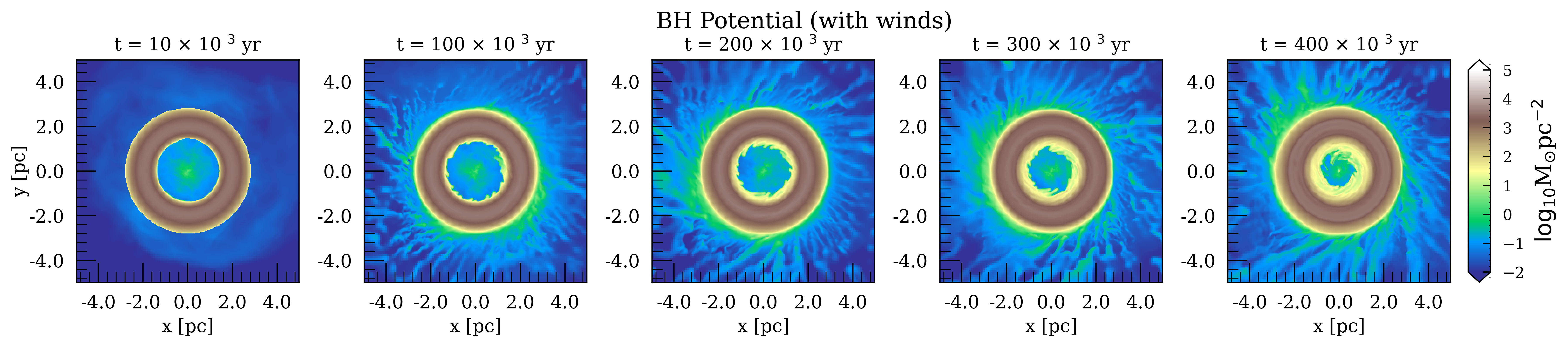}
    \includegraphics[width=0.95\textwidth]{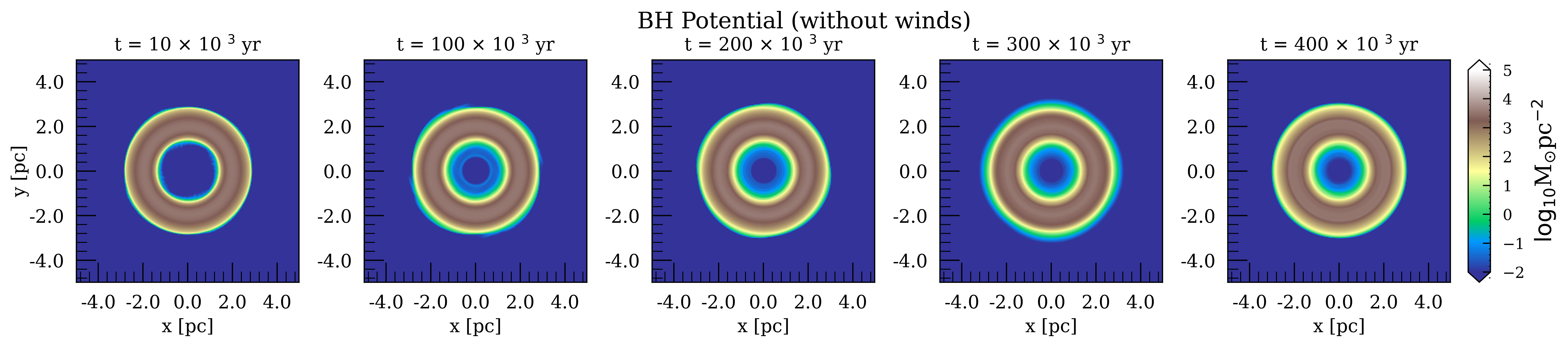}
    \caption{Different snapshots in time of the z-integrated density for the disk with (top) a black hole and nuclear star cluster potential with stellar winds, (middle) only the black hole potential with stellar winds and (bottom) a black hole potential without the stellar winds. These were taken at $t = 10 \times 10^3$, $100 \times 10^3$, $200 \times 10^3$, $300 \times 10^3$ and $400 \times 10^3$ years respectively. Note the inward migration of the inner edge of the disks as well as the formation of several transient filament like structures on the outer edge when the stellar winds are present (top and middle). The filaments form from the outward angular momentum transport of the in-falling gas. The bottom panel shows that hydrodynamic equilibrium is maintained without the presence of winds.}
    \label{fig:integrated_density}
\end{figure}

In this Appendix we compare simulations that include the Nuclear Star Cluster gravitational potential to those that include only the point source potential for the black hole. 
The initial conditions of the disks in the two potentials slightly differ in order to satisfy hydrodynamic equilibrium; the disk in the presence of the Nuclear Cluster potential has a mass and initial temperature of $3 \times 10^{4}$ M$_{\odot}$ and $4.5 \times 10^3$ K respectively, whereas the disk in just the BH potential has a mass and initial temperature of $4 \times 10^{4}$ M$_{\odot}$ and $10^3$ K respectively.
Figure \ref{fig:integrated_density} shows density integrated along its spin axis from the two different simulations. We also include a simulation with point source gravity and without the stellar winds in order to isolate their effects. The top row includes the gravitational potential from both Sgr A* and the NSC with stellar winds. The middle row includes the Sgr A* potential only with the stellar winds while the bottom row includes the Sgr A* potential without stellar winds.

The presence of the stellar winds clearly causes the instabilities seen in the inner edge as no such structures are seen in the simulation without the stellar winds. The accretion from the inner edge of the CND happens slightly faster in the case with NSC potential. This is because the matter on the inner edge of the CND loses the centrifugal support at the same rate but falls in faster with the increased gravitational potential.

\begin{figure}
    \centering
    \includegraphics[width=0.95\textwidth]{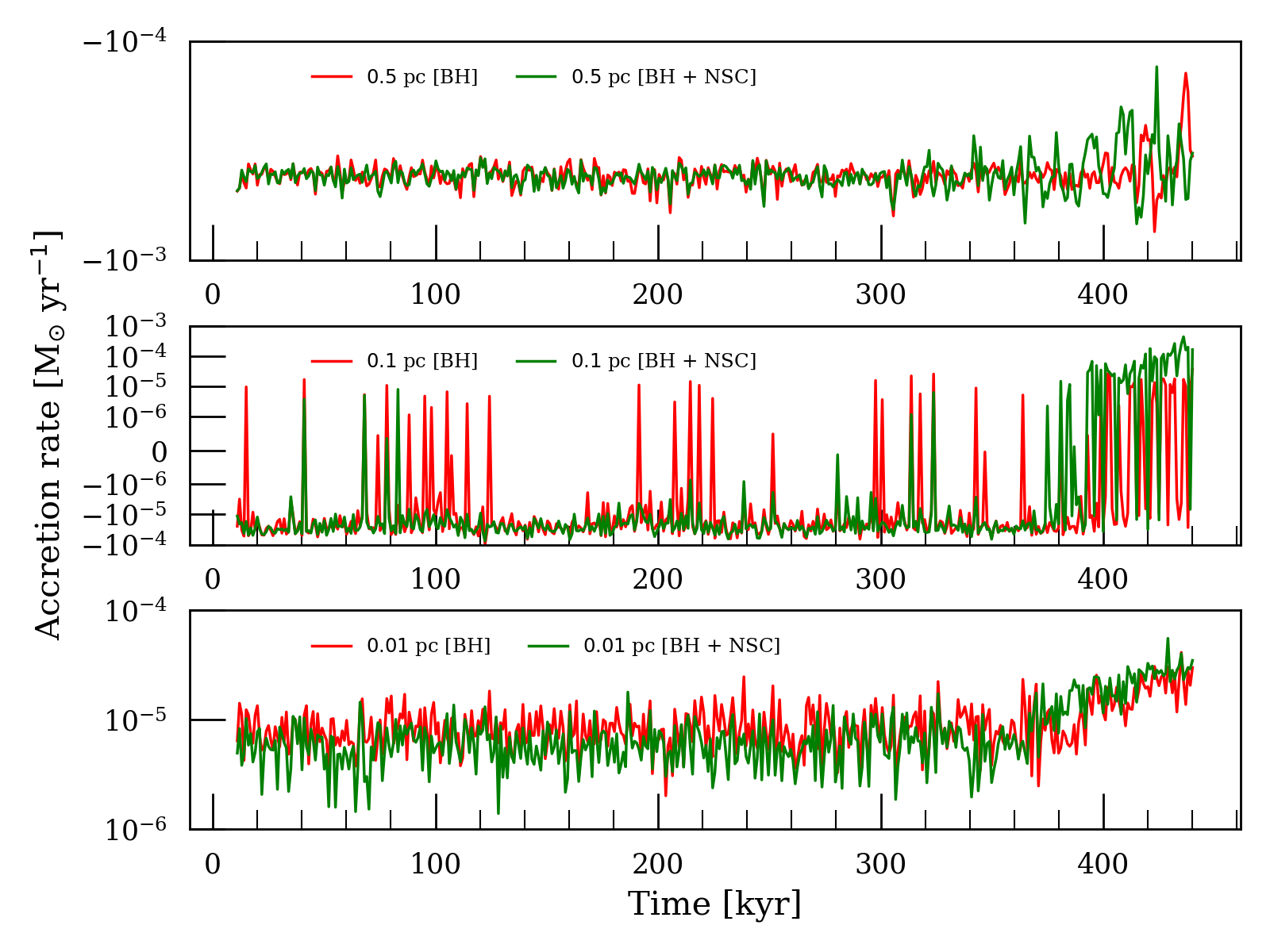}
    \caption{Comparison of the accretion rates between the simulations with different potentials. We compute the integrated accretion rates over a full solid angle at two different radii. The outflow at r$=0.5$ pc is close to equal in both the potentials because the NSC potential is very small at this radius. Similarly, the differences in accretion rate at r$=0.01$ pc are small.}
    \label{fig:accretion_rate_both}
\end{figure}

The accretion rates from the disks at $r=0.5$, $r=0.1$ and $0.01$ pc are plotted in Figure \ref{fig:accretion_rate_both}. The outflow at r$=0.5$ pc is close to equal in both the potentials because the winds from the stars have a super Keplerian velocity and are not strongly effected by the gravitational potential. Similarly the differences at r$=0.01$ pc are small because most of the inflowing matter is from the stellar winds and the NSC potential is very small at that radius.

\end{document}